# An ALMA Spectroscopic Survey of the Brightest Submillimeter Galaxies in the SCUBA-2-COSMOS Field (AS2COSPEC): Physical Properties of z = 2–5 Ultra- and Hyperluminous Infrared Galaxies

Cheng-Lin Liao (廖政霖)[1,2] , Chian-Chou Chen (陳建州)[1] , Wei-Hao Wang (王為豪)[1] , Ian Smail[3] , Y. Ao[4] ,
S. C. Chapman[5,6,7], U. Dudzevičiūtė[8], M. Frias Castillo[9], Minju M. Lee[10,11] , Stephen Serjeant[12] , A. M. Swinbank[3] ,
Dominic J. Taylor[3] , Hideki Umehata[13,14,15] , and Y. Zhao[16,17]

[1] Academia Sinica Institute of Astronomy and Astrophysics (ASIAA), No. 1, Sec. 4, Roosevelt Road, Taipei 10617, Taiwan; clliao@asiaa.sinica.edu.tw, ccchen@asiaa.sinica.edu.tw
[2] Graduate Institute of Astronomy, National Taiwan University, Taipei 10617, Taiwan
[3] Centre for Extragalactic Astronomy, Department of Physics, Durham University, South Road, Durham DH1 3LE, UK
[4] Purple Mountain Observatory and Key Laboratory for Radio Astronomy, Chinese Academy of Sciences, Nanjing, People's Republic of China
[5] Department of Physics and Atmospheric Science, Dalhousie University, Halifax, NS, B3H 4R2, Canada
[6] National Research Council, Herzberg Astronomy and Astrophysics, 5071 West Saanich Road, Victoria V9E 2E7, Canada
[7] Department of Physics and Astronomy, University of British Columbia, 6225 Agricultural Road, Vancouver V6T 1Z1, Canada
[8] Max-Planck-Institut für Astronomie, Königstuhl 17, D-69117 Heidelberg, Germany
[9] Leiden Observatory, Leiden University, P.O. Box 9513, 2300 RA Leiden, The Netherlands
[10] Cosmic Dawn Center (DAWN), Jagtvej 128, DK-2200 Copenhagen N, Denmark
[11] DTU-Space, Technical University of Denmark, Elektrovej 327, DK-2800 Kgs. Lyngby, Denmark
[12] School of Physical Sciences, The Open University, Walton Hall, Milton Keynes MK7 6AA, UK
[13] Institute for Advanced Research, Nagoya University, Furocho, Chikusa, Nagoya 464-8602, Japan
[14] Department of Physics, Graduate School of Science, Nagoya University, Furocho, Chikusa, Nagoya 464-8602, Japan
[15] Cahill Center for Astronomy and Astrophysics, California Institute of Technology, MS 249-17, Pasadena, CA 91125, USA
[16] Yunnan Observatories, Chinese Academy of Sciences, Kunming 650216, People's Republic of China
[17] Key Laboratory of Radio Astronomy and Technology Chinese Academy of Sciences, A20 Datun Road, Chaoyang District, Beijing, 100101, People's Republic of China

## Abstract

We report the physical properties of the 18 brightest ($S_{870\,\mu m} = 12.4$–19.2 mJy) and not strongly lensed 870 $\mu$m–selected dusty star-forming galaxies (DSFGs), also known as submillimeter galaxies (SMGs), in the COSMOS field. This sample is part of an ALMA band 3 spectroscopic survey (AS2COSPEC), and spectroscopic redshifts are measured in 17 of them at $z = 2$–5. We perform spectral energy distribution analyses and deduce a median total infrared luminosity of $L_{IR} = (1.3 \pm 0.1) \times 10^{13}$ $L_\odot$, infrared-based star formation rate (SFR) of SFR$_{IR}$ = $1390 \pm 150 M_\odot$ yr$^{-1}$, stellar mass of $M_* = (1.4 \pm 0.6) \times 10^{11} M_\odot$, dust mass of $M_{dust} = (3.7 \pm 0.5) \times 10^9 M_\odot$, and molecular gas mass of $M_{gas} = (\alpha_{CO}/0.8)(1.2 \pm 0.1) \times 10^{11} M_\odot$, suggesting that they are one of the most massive, ISM-enriched, and actively star-forming systems at $z = 2$–5. In addition, compared to less massive and less active galaxies at similar epochs, SMGs have comparable gas fractions; however, they have a much shorter depletion time, possibly caused by more active dynamical interactions. We determine a median dust emissivity index of $\beta = 2.1 \pm 0.1$ for our sample, and by combining our results with those from other DSFG samples, we find no correlation of $\beta$ with redshift or infrared luminosity, indicating similar dust grain compositions across cosmic time for infrared luminous galaxies. We also find that AS2COSPEC SMGs have one of the highest dust-to-stellar mass ratios, with a median of $0.02 \pm 0.01$, significantly higher than model predictions, possibly due to too-strong active galactic nucleus feedback implemented in the model. Finally, our complete and uniform survey enables us to put constraints on the most massive end of the dust and molecular gas mass functions.

*Unified Astronomy Thesaurus concepts:* Galaxy evolution (594); Galaxy formation (595); Infrared galaxies (790); Interstellar line emission (844); Interstellar medium (847); Submillimeter astronomy (1647)

## 1. Introduction

The population of submillimeter galaxies (SMGs; Smail et al. 1997; Barger et al. 1998; Hughes et al. 1998; Eales et al. 1999) was first discovered more than two decades ago by the Submillimeter Common User Bolometer Array (SCUBA) mounted on the single-dish James Clerk Maxwell Telescope at 850 $\mu$m. Subsequent observations using higher-resolution interferometers such as the Atacama Large Millimeter/submillimeter Array (ALMA), the NOrthern Extended Millimeter Array, and the Submillimeter Array, have revealed the nature of these dusty star-forming galaxies (DSFGs). Studies have found that SMGs are dust-rich (Swinbank et al. 2013; da Cunha et al. 2015; Donevski et al. 2020; Dudzevičiūtė et al. 2020; Pantoni et al. 2021; $M_{dust} \gtrsim 10^8 M_\odot$) and gas-rich (Bothwell et al. 2013; Birkin et al. 2021; $M_{gas} \gtrsim 10^{10} M_\odot$) galaxies with active star formation (Gullberg et al. 2019; Frias Castillo et al. 2023; SFR ~ 10–1000 $M_\odot$ yr$^{-1}$) and high dust attenuation (Dudzevičiūtė et al. 2020; Shim et al. 2022; $A_V \gtrsim 3$) that span a wide redshift range ($z \sim 1$–6) where the number density peaks at cosmic noon ($z \sim 2$–3; Chapman et al. 2005; Danielson et al. 2017; Chen et al. 2022b). SMGs typically exhibit infrared luminosities, integrated from 8 to 1000 $\mu$m in the rest frame, higher than $10^{12} L_\odot$, which result from the reprocessing of radiation by dust that absorbs the UV/optical light and reemits energy in the infrared, thereby usually classified as ultra- or hyperluminous infrared galaxies (ULIRGs[18] or HyLIRGs[19]).

The similarity between the redshift distribution of SMGs (Chapman et al. 2005; Wardlow et al. 2011; Simpson et al. 2014; Chen et al. 2016; Dudzevičiūtė et al. 2020) and that of the cosmic star formation density (Madau & Dickinson 2014; Zavala et al. 2021) suggests that SMGs are located at the key epochs of cosmic stellar mass growth. Indeed, several studies reveal that SMGs could account for ~20%–60% of the cosmic star formation density at $z \gtrsim 1$ (Barger et al. 2012; Swinbank et al. 2013; Cowie et al. 2017; Dudzevičiūtė et al. 2020). However, it is still an open question as to what triggers their extreme star formation. The high SFRs may be induced by major mergers of gas-rich galaxies, similar to ULIRGs in the local Universe (Sanders et al. 1988; Narayanan et al. 2009; Chen et al. 2015; Perry et al. 2023). However, ample gas accretion from the intergalactic medium may trigger disk instability (Narayanan et al. 2015; Tacconi et al. 2020). One way to address this issue is to measure scaling relations with statistical samples of SMGs for gas depletion time and gas fractions against redshift and offset from the star formation main sequence (MS) and compare the results to those based on less active and less massive star-forming galaxies (Genzel et al. 2015; Tacconi et al. 2020), where such correlations have been explained through Toomre stability criteria (Toomre 1964) by Tacconi et al. (2020).

In addition to the star formation triggering mechanisms, understanding the role of dust is crucial for comprehending galaxy formation, particularly since the properties of dust-rich SMGs by observations are poorly reproduced by state-of-the-art simulations (Popping et al. 2017; Hou et al. 2019; Li et al. 2019; McAlpine et al. 2019). A commonly used method for studying the dust properties is fitting the far-infrared (FIR) spectral energy distribution (SED) with a modified blackbody (MBB) model, via which characteristic dust temperature, dust mass, and dust emissivity index can be estimated. For studies that lack sufficient photometry, $\beta$, which is degenerate with temperature, is commonly fixed. While $\beta$ is typically assumed to be between 1.5 and 2.0 (Scoville et al. 2017; Kaasinen et al. 2019; Dudzevičiūtė et al. 2020), recent studies suggest that $\beta$ is consistent with or possibly greater than 2 (Magnelli et al. 2012; Casey et al. 2021; da Cunha et al. 2021; Cooper et al. 2022; Bendo et al. 2023; McKay et al. 2023), which is similar to the value in the local Universe (Smith et al. 2013) and suggests a dust assembly scenario by dust grain growth in the interstellar medium (ISM). A detailed investigation of $\beta$ is useful not only in constraining dust grain growth models (Hirashita et al. 2014) but also in analyses of even higher-redshift galaxies that usually have sparse sampling of millimeter photometry.

Finally, an effective constraint for testing theoretical models is the mass functions, which represent the number density of objects within a given mass range and a redshift bin. This comparison can help us understand how different mass budgets assemble over cosmic time. Recent interferometric blind spectroscopic surveys conducted by ALMA and the Very Large Telescope (VLA), such as the ASPECS (Decarli et al. 2019) and the VLA COLDz survey (Riechers et al. 2019), have presented the evolution of the CO luminosity function, which is analogous to the gas mass function when assuming a constant $\alpha_{CO}$, from $z \sim 6$ to $z = 0$. While effective, however, due to a small field of view for interferometers, the constraints at the massive end of the gas mass functions are limited. A semianalytical model developed by Béthermin et al. (2022) predicts that the $L'_{CO}$ function can reach to the brightest end at ~2 × $10^{11} L_\odot$, successfully describing the observational results up to this limit. Observational measurements of dust mass functions have been presented at epochs up to $z \approx 3$ (Magnelli et al. 2019; Pozzi et al. 2019). However, state-of-the-art simulations (Popping et al. 2017; Hou et al. 2019; Li et al. 2019; McAlpine et al. 2019) have struggled to reproduce the observed high number density of high dust mass sources, highlighting the need for a different treatment of dust production in the model but also more measurements at the massive end to better constrain the dust production of early massive galaxies.

Recently, a catalog of approximately 1000 sources with 850 $\mu$m flux ($S_{850}$) ranging from 2 to 20 mJy was presented by Simpson et al. (2019) from the SCUBA-2 Cosmology Legacy Survey in COSMOS (S2COSMOS), covering an area of 1.6 deg². Follow-up observations of the 182 brightest S2COSMOS sources ($S_{850} > 6.2$ mJy) were carried out using the ALMA band 7 continuum, and the results were presented as the AS2COSMOS sample by Simpson et al. (2020), where they deblended the single-dish sources into 260 SMGs with 870 $\mu$m flux ($S_{870}$) ranging from 0.7 to 19.2 mJy with precise localization. To further exploit the complete selection of the brightest SMGs in the AS2COSMOS sample, we have recently carried out an ALMA band 3 spectroscopic survey of the 18 brightest AS2COSMOS SMGs. The first results regarding line detection of CO and [C I], redshift distributions, and their lensing nature have been presented in Chen et al. (2022b). In this study, we perform detailed analyses of the CO and [C I] emission lines, which are tracers of molecular gas, and conduct SED fittings of the X-ray–to–radio photometry to further estimate other physical properties of the AS2COSPEC sample.

The paper is structured as follows. In Section 2, we provide an overview of the AS2COSPEC sample and describe the ancillary multiwavelength photometry. In Section 3, we present analyses to estimate the physical properties of AS2COSPEC SMGs, including SED fittings using Code Investigating GALaxy Emission (X-CIGALE) and MBB models. In Section 4, we discuss various aspects of these properties, including active galactic nucleus (AGN) fraction, gas depletion time, dust emissivity index, and gas and dust mass functions. Finally, in Section 5, we summarize our results. We adopt a flat $\Lambda$CDM cosmological model with $H_0 = 67.7$ km s⁻¹ Mpc⁻¹, $\Omega_M = 0.31$, and $\Omega_\Lambda = 0.69$ (Planck Collaboration et al. 2020) throughout this paper.

## 2. Observation and Data

### 2.1. Sample and ALMA Band 3 Data

The AS2COSPEC sample presented in this work represents the 18 brightest ($S_{870} = 12.4$–19.2 mJy) SMGs drawn from the parent sample of 260 AS2COSMOS SMGs. These were originally detected by the ALMA band 7 follow-up continuum observations of the submillimeter sources uncovered by the SCUBA-2 survey at 850 $\mu$m in the COSMOS

---

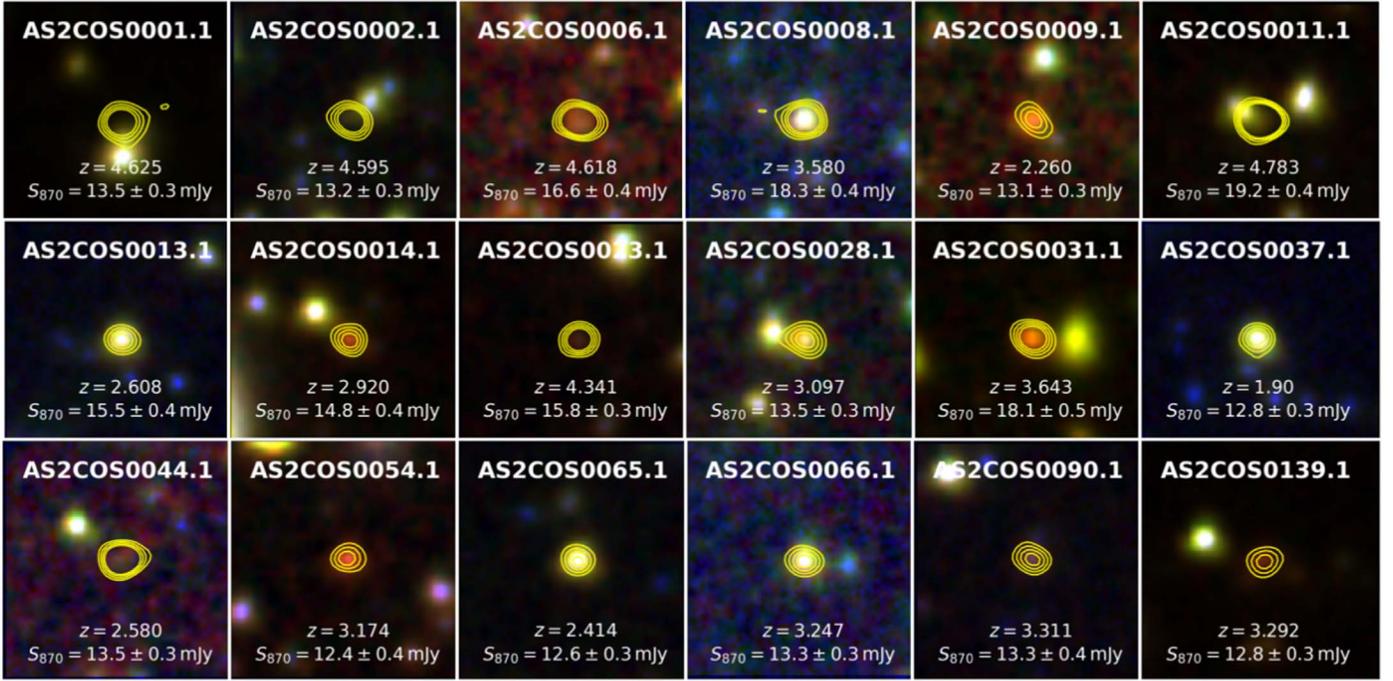

**Figure 1.** Thumbnail images of our sample SMGs sized $20'' \times 20''$, where the background color map uses IRAC 4.5 $\mu$m, IRAC 3.6 $\mu$m, and UVISTA $K_s$ bands as red, green, and blue, respectively. The contours show the ALMA band 3 continuum at the $5\sigma$, $7\sigma$, $9\sigma$, and $11\sigma$ levels. The spectroscopic redshift (Chen et al. 2022b) and 870 $\mu$m flux density (Simpson et al. 2020) of each source are also labeled on the image. We note that some of the sources have very weak detection or even no detection in the near-infrared due to dust extinction, but they are robustly detected in the FIR.

field (Simpson et al. 2019, 2020). Data and data reduction of our ALMA band 3 spectral scan observations were presented in detail in Chen et al. (2022b). Here, we briefly describe the relevant points.

All 18 AS2COSPEC SMGs are detected in the band 3 (3 mm) continuum with high significance (signal-to-noise ratio, S/N $\gtrsim$ 10) in the naturally weighted maps (Figure 1). Given the typical resolution of $4''$, the 3 mm continuum is found to be not spatially resolved except for the three sources that have close companions, AS2COS0001.1, AS2COS0008.1, and AS2COS0028.1. For these three sources, their continuum images are produced with Briggs weighting, resulting in sufficient spatial resolutions for separating the main source and its companion. The 3 mm photometry is simply deduced from the IMFIT task in CASA, which yields consistent results to estimates based on fittings in the visibility space. The resulting flux densities are provided in Table 1.

Spectra are first extracted using an aperture equal to the synthesized beam, and the line intensities are measured using Gaussian profile modeling. Corrections to the total line intensities are then applied based on the curve-of-growth analyses. Line emission is detected in all but one AS2COSPEC SMG, and their line intensities, line widths, and corresponding luminosities are measured via line moments (Chen et al. 2022b). The results are provided in Table 1.

Spectroscopic redshifts are determined based on the measured emission lines. The majority (14 out of 18) of the sample SMGs have multiple emission lines from the original band 3 data, as well as other follow-up observations (Mitsuhashi et al. 2021; Frias Castillo et al. 2023; D. J. Taylor et al. 2023, in preparation). Three SMGs with single-line detection (AS2COS0044.1, AS2COS0066.1, and AS2COS0090.1) have their redshifts assigned to one that is closest to their photometric redshifts. Estimates based on sources with multiple line detections suggest that such an assignment of redshifts for single-line sources should be correct in $\sim$80% of the cases. For the remaining $\sim$20% of the cases, their photometric redshifts do not accurately capture the spectroscopic values ($|z_{phot} - z_{spec}| > 1$). The reason for this mismatch is currently unclear, possibly due to the fact that the modeling of the photometric redshifts does not fully capture the properties of this heavily obscured population. The only source that does not have any line detection (AS2COS0037.1) is assumed to be at the redshift gap ($z = 1.74$–2.05) due to the frequency coverage of the band 3 scan. Although our sample of SMGs is considerably brighter than the typical ones, strong gravitational lensing with magnification of greater than 2 can only be found in at most one source (AS2COS0002.1; Table 1). Since the evidence for lensing is weak, throughout this paper, we simply ignore lensing effects. We confirm, however, that including lensing corrections or not would not change our results significantly.

### 2.2. Multiwavelength Data

The COSMOS field (Scoville et al. 2007) has been a prime target field covering a wide wavelength range from X-ray to radio. We exploit these exquisite data sets and use them to measure and model the SED of our sample SMGs, which allows us to obtain estimates of their physical properties, such as stellar masses and SFRs. In Figure 1, we show postage stamps of all 18 sample SMGs using infrared imaging and overlay the 3 mm continuum from ALMA observations in contours. For our work, we mainly select and adopt the multiwavelength photometry collected by Simpson et al. (2020) for the parent AS2COSMOS sample. In the following,

**Table 1**
ALMA Band 3 Line and Continuum Measurements

| Source | Line | Redshift | $I_\nu$ (Jy km s$^{-1}$) | FWHM (km s$^{-1}$) | $L'_{\mathrm{CO},J/[\mathrm{C\,II}](1-0)}$ (10$^{10}$ K km s$^{-1}$ pc$^2$) | $S_{3\,\mathrm{mm}}$ (mJy) | $\mu$ | Other Lines |
|---|---|---|---|---|---|---|---|---|
| AS2COS0001.1 | CO(5–4) | 4.6237 ± 0.0007 | 0.8 ± 0.2 | 770 ± 110 | 2.7 ± 0.6 | 0.29 ± 0.03 | 1.4$^{+0.1}_{-0.1}$ | [C II][a] |
| AS2COS0002.1 | CO(5–4) | 4.5956 ± 0.0006 | 1.2 ± 0.3 | 600 ± 50 | 3.8 ± 0.8 | 0.27 ± 0.02 | 3.0$^{+1.4}_{-0.7}$ | [C II][a] |
| AS2COS0006.1 | CO(5–4) | 4.6183 ± 0.0001 | 2.6 ± 0.3 | 930 ± 6 | 8.5 ± 1.1 | 0.34 ± 0.02 | 1.0 | [C II][a] |
| AS2COS0008.1 | CO(4–3) | 3.5811 ± 0.0004 | 1.5 ± 0.2 | 610 ± 30 | 5.2 ± 0.6 | 0.28 ± 0.02 | 1.0 | CO(1–0)[b] |
| AS2COS0009.1 | CO(3–2) | 2.2599 ± 0.0002 | 2.9 ± 0.4 | 610 ± 40 | 8.4 ± 1.2 | 0.18 ± 0.02 | 1.0 | CO(6–5)[d] |
| AS2COS0011.1 | [C I](1–0) | 4.7831 ± 0.0007 | 1.1 ± 0.3 | 640 ± 60 | 5.1 ± 1.5 | 0.47 ± 0.02 | 1.0 | ⋯ |
| | CO(5–4) | 4.7830 ± 0.0002 | 1.9 ± 0.2 | 530 ± 20 | 6.7 ± 0.8 | | | |
| AS2COS0013.1 | CO(3–2) | 2.6079 ± 0.0001 | 3.5 ± 0.2 | 320 ± 20 | 13.0 ± 0.8 | 0.19 ± 0.03 | 1.0 | CO(1–0)[b] |
| AS2COS0014.1 | CO(3–2) | 2.9202 ± 0.0005 | 2.1 ± 0.3 | 720 ± 30 | 9.1 ± 1.5 | 0.17 ± 0.02 | 1.7$^{+0.5}_{-0.2}$ | CO(6–5)[d] |
| AS2COS0023.1 | CO(4–3) | 4.3410 ± 0.0001 | 1.9 ± 0.1 | 420 ± 20 | 9.1 ± 0.7 | 0.22 ± 0.02 | 1.7$^{+0.1}_{-0.1}$ | CO(1–0)[b] |
| | CO(5–4) | 4.3414 ± 0.0002 | 1.8 ± 0.2 | 390 ± 30 | 5.5 ± 0.6 | | | |
| AS2COS0028.1 | CO(3–2) | 3.0966 ± 0.0003 | 0.9 ± 0.2 | 640 ± 60 | 4.5 ± 0.8 | 0.21 ± 0.03 | 1.0 | ⋯ |
| | CO(4–3) | 3.0964 ± 0.0002 | 1.6 ± 0.2 | 700 ± 40 | 4.3 ± 0.6 | | | |
| AS2COS0031.1 | CO(4–3) | 3.6432 ± 0.0001 | 2.5 ± 0.2 | 410 ± 10 | 8.9 ± 0.7 | 0.28 ± 0.02 | 1.0 | CO(1–0)[b] |
| | [C I](1–0) | 3.6431 ± 0.0004 | 1.0 ± 0.2 | 400 ± 60 | 3.1 ± 0.7 | | | |
| AS2COS0037.1 | ⋯ | 1.90 ± 0.05 | ⋯ | ⋯ | ⋯ | 0.15 ± 0.02 | 1.0 | ⋯ |
| AS2COS0044.1 | CO(3–2) | 2.5793 ± 0.0002 | 1.4 ± 0.2 | 690 ± 30 | 5.0 ± 0.6 | 0.33 ± 0.02 | 1.0 | ⋯ |
| AS2COS0054.1 | CO(4–3) | 3.1735 ± 0.0004 | 1.5 ± 0.3 | 720 ± 80 | 4.4 ± 1.0 | 0.15 ± 0.02 | 1.0 | CO(1–0)[b] |
| AS2COS0065.1 | CO(3–2) | 2.4140 ± 0.0002 | 2.5 ± 0.2 | 510 ± 10 | 8.1 ± 0.8 | 0.15 ± 0.02 | 1.0 | CO(6–5)[d] |
| AS2COS0066.1 | CO(4–3) | 3.2492 ± 0.0005 | 1.6 ± 0.3 | 640 ± 70 | 4.8 ± 0.9 | 0.17 ± 0.02 | 1.0 | ⋯ |
| AS2COS0090.1 | CO(4–3) | 3.3137 ± 0.0003 | 1.3 ± 0.3 | 760 ± 80 | 3.9 ± 0.8 | 0.14 ± 0.02 | 1.0 | ⋯ |
| AS2COS0139.1 | CO(4–3) | 3.2923 ± 0.0002 | 2.6 ± 0.3 | 380 ± 30 | 7.9 ± 0.8 | 0.17 ± 0.03 | 1.0 | Lyα[c] and CO(8–7)[d] |
| Mean | ⋯ | 3.5695 ± 0.0001 | 1.8 ± 0.1 | 590 ± 10 | 6.3 ± 0.2 | 0.23 ± 0.01 | 1.21$^{+0.08}_{-0.04}$ | ⋯ |
| Median | ⋯ | 3.3137 ± 0.2857 | 1.6 ± 0.2 | 610 ± 50 | 5.0 ± 1.1 | 0.20 ± 0.03 | 1.00 ± 0.03 | ⋯ |

**Notes.** AS2COS0011.1, AS2COS0023.1, AS2COS0028.1, and AS2COS0031.1 have double line detections in our ALMA band 3 spectroscopic survey, and AS2COS0037.1 does not have any line detection. The luminosity presented here is the line luminosity of the corresponding quantum transition. Lensing magnifications ($\mu$) are obtained from Chen et al. (2022b), where uncertainties less than 0.05 are omitted.
[a] Mitsuhashi et al. (2021).
[b] Frias Castillo et al. (2023).
[c] D. J. Taylor et al. (2023, in preparation).
[d] Daddi et al. (2022).

we briefly summarize their analyses and refer to Simpson et al. (2020) for more details.

The photometry measurements from Simpson et al. (2020) can be summarized as follows. First, the 260 AS2COSMOS SMGs were cross-matched to the COSMOS2015 catalog (Laigle et al. 2016) with a chosen matching radius of 0″85, which was shown via simulations to yield a false match rate of ∼7%. Second, Simpson et al. (2020) also included a few broad bands that had deeper images compared to those included in the COSMOS2015 catalog, so the photometry of those bands was updated. In particular, the $YJHK_s$ photometry was extracted based on the fourth data release of the UltraVISTA survey (McCracken et al. 2012), instead of the second release that was used in the COSMOS2015 catalog. In addition, the photometry of the $grizY$ bands was also replaced by the catalog provided by the Hyper Suprime-Cam (HSC) Subaru Strategic Program (SSP; Aihara et al. 2018). Necessary aperture-to-total corrections were applied in order to obtain total flux densities for these bands. Duplicated observations were taken by different instruments for a few common filters, such as the $r$ and $i$ bands from HSC-SSP and Suprime-Cam, as well as the $H$ and $K_s$ bands from VIRCAM and WIRCAM. In general, we adopt the deeper measurements if multiple choices are available. The 260 AS2COSMOS SMGs were also cross-matched to the Chandra COSMOS-Legacy Survey (Civano et al. 2016) within the 3σ positional uncertainty of the X-ray location. Finally, we only use the broadband measurements, since most of our SMGs are not detected in the narrow and medium bands, and including this photometry or not does not affect the results significantly.

In total, four of our 18 SMGs (AS2COS0014.1, AS2COS0028.1, AS2COS0031.1, and AS2COS0066.1) can be matched to the catalog of the Chandra COSMOS-Legacy Survey (Civano et al. 2016), in which they employed the maximum-likelihood algorithm for source extraction. Two of our 18 SMGs (AS2COS0037.1 and AS2COS0065.1) have robust detections (S/N ⩾ 3) in the $u$-band data from the Canada–France–Hawaii Telescope (CFHT). Approximately 60% of the sources are detected in the optical bands ($B$, $V$, $g$, $r$, $i$, $z^+$) from Subaru Suprime-Cam and HSC, and about 70% of the sources are detected in $Y$, $J$, $H$, and $K_s$ from the DR4 UltraVISTA survey.

For mid-infrared and FIR photometry, various deblending techniques were employed to address the issue of source confusion, which was caused by the modest spatial resolution of the corresponding images. For images taken by the Infrared Array Camera (IRAC; Fazio et al. 2004) on the Spitzer Space Telescope (hereafter Spitzer), IRACLEAN (Hsieh et al. 2012) was used for deblending, with an updated prior catalog by including the AS2COSMOS SMGs in the original prior made from a stacked $zYHK_s$ image. For the 24 μm image taken by the Multiband Imaging Photometer (MIPS; Rieke et al. 2004) on Spitzer, as well as the 100 and 160 $\mu$m images taken by the Photodetector Array Camera and Spectrometer (PACS; Poglitsch et al. 2010) on the Herschel Space Telescope (hereafter Herschel), the "super-deblended" photometry was adopted, which was made by deblending those images with a $K_s+$ radio prior (Jin et al. 2018). Finally, for the images at 250, 350, and 500 $\mu$m taken by the SPIRE instrument on Herschel, deblending was performed following the procedures described by Swinbank et al. (2013), where they adopted prior combining sources detected at 24 $\mu$m, VLA 3 GHz, and ALMA 870 $\mu$m.

A good fraction of the sample was targeted by previous ALMA and/or Plateau de Bure Interferometer observations at 1.2–1.3 mm (Brisbin et al. 2017; Smolčić et al. 2017), and the photometry is included in Simpson et al. (2020). Finally, for radio, the 3 GHz photometry was obtained by again cross-matching the AS2COSMOS SMGs to the catalog of the VLA-COSMOS 3 GHz Large Project (Smolčić et al. 2017). For the 3 GHz faint sources, the flux densities are deduced by taking the pixel values at the positions of the SMGs, which were then corrected to total flux densities based on morphological analyses of the stacked images. In addition to the 3 GHz photometry obtained by Simpson et al. (2020), we further cross-match AS2COSMOS SMGs to the VLA-COSMOS 1.4 GHz catalog presented by Schinnerer et al. (2010) using a matching radius of 1″. We identify 95 1.4 GHz counterparts among the 260 AS2COSMOS SMGs, and 10 of them are our AS2COSPEC SMGs. For the remaining eight SMGs in our sample that are not detected, we determine their flux density upper limit as 48 $\mu$Jy, which is the detection limit of the VLA-COSMOS Deep Project.

Overall, the number of AS2COSPEC SMGs that is detected in each of these bands described above ranges from 1 to 18. In Table 2, we provide a summary of the number of detections for each band.

## 3. Analyses and Results

### 3.1. X-CIGALE SED Fitting

We employ the 2022.1 version of X-CIGALE[20] (Boquien et al. 2019; Yang et al. 2020, 2022) to model the SED of our SMGs. Briefly, X-CIGALE is a package written in Python that aims at fitting multiwavelength photometry, from X-ray to radio, with the goal of estimating the physical properties of galaxies by the method of Bayesian analysis. Several emitting sources are taken into account in the fitting, including stellar emission, ionized nebular emission, AGN emission dust emission, and synchrotron emission. In Table 3, we present the selected models and the chosen range of parameters. We use the photometry described in Section 2 and summarized in Table 2 in the fitting. Any of the photometry that has an S/N lower than 3 is treated as an upper limit in X-CIGALE.

The basic principle that we follow is simply to assemble a set of parameter space enabled by X-CIGALE that produces the best fits with the lowest reduced $\chi^2$ across the whole sample. As a check, we compare our results to those obtained from fittings using MAGPHYS, following the procedures presented in Dudzevičiūtė et al. (2020), where they established their methods based on testing against simulated galaxies. In the following, we provide justifications and highlight a few

**Table 2**
Photometry Used in the SED Fitting and the Detection Situation in Each Band

| Instrument/Telescope | Filter | S/N ⩾ 3 | S/N < 3 | No Coverage |
|---|---|---|---|---|
| Chandra | Soft | 1 | 17 | 0 |
| | Hard | 0 | 18 | 0 |
| | Full | 1 | 17 | 0 |
| MegaCam/CFHT | $u$ | 2 | 12 | 4 |
| Suprime-Cam/Subaru | $B$ | 8 | 6 | 4 |
| | $V$ | 9 | 5 | 4 |
| | $r$ | 13 | 1 | 4 |
| | $i$ | 13 | 1 | 4 |
| | $z^{++}$ | 10 | 4 | 4 |
| HSC/Subaru | $g$ | 11 | 0 | 7 |
| | $r$ | 11 | 0 | 7 |
| | $i$ | 11 | 0 | 7 |
| | $z$ | 11 | 0 | 7 |
| | $y$ | 11 | 3 | 4 |
| VIRCAM/VISTA | $Y$ | 11 | 4 | 3 |
| | $J$ | 11 | 4 | 3 |
| | $H$ | 13 | 2 | 3 |
| | $K_s$ | 14 | 1 | 3 |
| WIRCAM/CFHT | $H$ | 5 | 9 | 4 |
| | $K_s$ | 7 | 7 | 4 |
| IRAC/Spitzer | 3.6 $\mu$m | 18 | 0 | 0 |
| | 4.5 $\mu$m | 18 | 0 | 0 |
| | 5.8 $\mu$m | 16 | 2 | 0 |
| | 8.0 $\mu$m | 17 | 1 | 0 |
| MIPS/Spitzer | 24 $\mu$m | 13 | 5 | 0 |
| PACS/Herschel | 100 $\mu$m | 2 | 15 | 1 |
| | 160 $\mu$m | 6 | 11 | 1 |
| SPIRE/Herschel | 250 $\mu$m | 13 | 0 | 5 |
| | 350 $\mu$m | 15 | 1 | 2 |
| | 500 $\mu$m | 14 | 1 | 3 |
| ALMA | 870 $\mu$m | 18 | 0 | 0 |
| | 1200 $\mu$m | 10 | 0 | 8 |
| | 3000 $\mu$m | 18 | 0 | 0 |
| VLA | 3 GHz | 18 | 0 | 0 |
| | 1.4 GHz | 10 | 0 | 8 |

**Note.** The extraction and the estimation of the photometry in each band is described in Section 2.2.

findings on a number of key selections. We present the testing results at the end of this section.

#### 3.1.1. Star Formation History

We adopt a delayed exponential star formation history (SFH) model with an exponential burst. This assumption of two independent components of SFH is now commonly assumed, which is also adopted in MAGPHYS, and it has been argued that it better reproduces the stellar masses of SMGs (Michałowski et al. 2014). By experimenting with different ranges of the input parameters, we find that including a starburst (SB) as recent as 10 Myr is needed for more than half of the sample SMGs, which enables significantly better fitting results on FIR photometry. On average, compared to the fitting without this 10 Myr burst included, that having 10 Myr in the

**Table 3**
The Selected Modules and Input Parameters in CIGALE

| Modules | Models | Parameters | Values |
|---------|--------|------------|--------|
| SFH | $SFR(t) = SFR_{delayed}(t) + SFR_{burst}(t)$ | tau_main [Myr] | 100, 500, 1000, 5000 |
| | | age_main [Myr] | 500, 1000 |
| | | tau_burst [Myr] | 50, 100, 150 |
| | | age_burst [Myr] | 10, 100, 300 |
| | | f_burst | 0.1, 0.5, 0.9 |
| Simple stellar population | bc03 (Bruzual & Charlot 2003) | IMF | 1 (Chabrier 2003) |
| Nebula | Continuum and line nebular emission | f_dust | 0.7 |
| | | lines_width [km s$^{-1}$] | 500 |
| Dust attenuation | CF00 (Charlot & Fall 2000) | Av_ISM | 0.3, 0.8, 1.2, 3.3, 3.8 |
| | | mu | 0.44, 0.01, 0.001 |
| | | filters | SUBARU_V |
| Dust emission | dl2014 (Draine et al. 2014) | qpah | 0.47, 3.90, 7.32 |
| | | umin | 10.0, 30.0, 50.0 |
| | | alpha | 2.0, 2.5, 3.0 |
| AGN | skirtor2016 (Stalevski et al. 2012, 2016) | i | 10, 40, 80 |
| | | fracAGN | 0.01, 0.05, 0.10, 0.30, 0.50 |
| X-ray | From AGN and galaxy | *default* | *default* |
| Radio | Synchrotron emission | qir_sf | 1.0, 2.0, 3.0 |
| | | alpha_sf | see Table 4 |
| | | R_agn | 0.1, 1.0, 10.0 |
| | | alpha_agn | see Table 4 |

**Note.** Parameters that are not listed in this table are the X-CIGALE default values. The alpha_sf and alpha_agn values in the radio module vary from source to source, which depends on the radio synchrotron slope of the sources. See Section 3.1.7 and Table 4 for more details.

SB model decreases the reduced $\chi^2$ by 20% and boosts the $L_{IR}$ by 90%.

### 3.1.2. Simple Stellar Population

We employ the stellar population model from Bruzual & Charlot (2003) with a Chabrier initial mass function (IMF). The stellar metallicity is set to the solar value. We note that some recent studies have found evidence in favor of a non-Chabrier IMF (Zhang et al. 2018; Cai et al. 2020), which, however, has not been confirmed by other studies (Lagos et al. 2020; Lovell et al. 2021). Since this is still a topic of debate, we choose to adopt the Chabrier IMF, which has also been adopted by recent studies of DSFGs (Donevski et al. 2020; Dudzevičiūtė et al. 2020; Cardona-Torres et al. 2023).

### 3.1.3. Nebular Emission

We adopt nebular emission in X-CIGALE, which is based on the template from Inoue (2011). This includes hydrogen continuum emission and line emission from He II at 30.38 nm to [N II] at 205.4 $\mu$m.

### 3.1.4. Dust Attenuation

We adopt the two-component dust model (Charlot & Fall 2000, CF00) as the attenuation model in our SED fitting. This is motivated by the recent discoveries that the optical/ near-infrared and FIR emissions of SMGs are not necessarily colocated (Chen et al. 2015; Hodge et al. 2016; Smail et al. 2023). We set a stronger attenuation strength for the birth clouds than ISM, with slopes of $-1.3$ and $-0.7$ for the birth clouds and ISM, respectively. They are linked by the $\mu$ parameter, ranging from 0.001 to 0.44 of $A_{V,ISM} = 0.3$–3.8. We also repeat the SED fitting by adopting the SB attenuation law from Calzetti et al. (2000) but do not find overall better results based on reduced $\chi^2$.

### 3.1.5. Dust Emission

We adopt the dust emission model from Draine et al. (2014). Briefly, this model separates the dust emission into two components: the first one is the dust emission from the stellar light absorption of the general stellar population with a single radiation field $U_{min}$; the second one is the dust emission from the star-forming region with the radiation field ranging from $U_{min}$ to $U_{max}$ following the power-law index $\alpha$. There are four free parameters in this model: $U_{min}$, $\alpha$, $q_{PAH}$, and $\gamma$. $q_{PAH}$ is the mass fraction of polycyclic aromatic hydrocarbon (PAH) of the total dust mass, which we set to $q_{PAH} = 0.47$, 3.90, and 7.32, ranging from the minimum to maximum acceptable values in X-CIGALE. We set a wide tolerance of $q_{PAH}$ due to the fact that the observed 24 $\mu$m is located at the redshifted PAH emission region so that the 24 $\mu$m is sensitive to the PAH emission strength. Recent results from simulations also suggest a wide range of $q_{PAH}$ (Narayanan 2023).

### 3.1.6. AGN

AGN are thought to exist in some SMGs (Alexander et al. 2005) and could play a significant role in galaxy evolution. We adopt the AGN template from Stalevski et al. (2012, 2016), which models the central source emission in the torus with most of the dust in high-density clumps by considering the 3D radiative transfer. It also models the AGN emission from different lines of sight and derives the luminosity contribution

**Table 4**
Radio Spectral Index

| Source | $\alpha$ |
| --- | --- |
| AS2COS0001.1 | $2.1 \pm 0.3$ |
| AS2COS0002.1 | 0.7 |
| AS2COS0006.1 | $1.1 \pm 0.3$ |
| AS2COS0008.1 | 0.7 |
| AS2COS0009.1 | $0.8 \pm 0.2$ |
| AS2COS0011.1 | 0.7 |
| AS2COS0013.1 | $0.9 \pm 0.2$ |
| AS2COS0014.1 | 0.7 |
| AS2COS0023.1 | 0.7 |
| AS2COS0028.1 | $2.1 \pm 0.3$ |
| AS2COS0031.1 | 0.7 |
| AS2COS0037.1 | $0.4 \pm 0.2$ |
| AS2COS0044.1 | 0.7 |
| AS2COS0054.1 | $0.7 \pm 0.3$ |
| AS2COS0065.1 | $1.0 \pm 0.1$ |
| AS2COS0066.1 | $1.5 \pm 0.3$ |
| AS2COS0090.1 | 0.7 |
| AS2COS0139.1 | $0.8 \pm 0.2$ |

**Note.** Sources showing uncertainties are SMGs having both the 3 and 1.4 GHz photometry, so we derive their slopes from these two photometries. Those showing single photometry without uncertainties are SMGs having only the 3 GHz photometry, so we assume the slope of 0.7 (Smolčić et al. 2017) in the X-CIGALE SED fitting.

from AGN to the total dust luminosity (fracAGN). We set a wide range of input fracAGN, 0.01, 0.05, 0.10, 0.30, and 0.50, to let the model select the best value. We find that all sources prefer the fracAGN solution between 0.01 and 0.10 (see Table 5), which means that AGN play a minor role of contributing to the infrared emission from SMGs.

### 3.1.7. Radio

The model for radio emission is comprised of a simple power-law synchrotron emission with four parameters. Two of them are the slopes of the power-law synchrotron emission for star formation (alpha_sf) and AGN (alpha_agn); one is the value of the FIR/radio correlation coefficient for star formation (qir_sf), which is

$$\mathrm{qir\_sf} = \log\left(\frac{L_{\mathrm{IR}}}{3.75 \times 10^{12}\,\mathrm{W}}\right) - \log\left(\frac{L_{1.4\,\mathrm{GHz}}}{\mathrm{W\,Hz}^{-1}}\right); \qquad (1)$$

and the last one is the radio-loudness parameter for AGN (R_agn). We set a range of qir_sf and R_agn for X-CIGALE to find the best solution, and alpha_sf and alpha_agn are assumed to be the same. Ten of the SMGs in our sample have both 3 and 1.4 GHz photometry measurements; thus, their spectral slopes are measured individually. For the remaining eight SMGs that only have 3 GHz photometry, we assume a slope of 0.7, following the original study of the 3 GHz catalog (Smolčić et al. 2017). Table 4 summarizes the adopted values for alpha_sf and alpha_agn of each source, and we adopt these values as the input parameters in the radio model.

### 3.1.8. Fitting Results and Comparisons with MAGPHYS

The left panel of Figure 2 shows an example SED fit of one of our SMGs. The SED fitting results of all the sources are presented in Appendix B. Each physical model is plotted with a different colored line, and the total SED fitting model is the black curve, which is the summation of all the solid lines. The relative residual ((Observation − Model)/Observation) of each photometry is shown in the lower plot. Table 5 presents the physical properties of each source, and the statistical values (mean and median) are also listed. The physical properties listed in Table 5 are the stellar mass ($M_*$), $V$-band attenuation ($A_V$), total infrared luminosity ($L_{\mathrm{IR}}$), fraction of AGN luminosity to total dust luminosity (fracAGN), SFR, and dust mass ($M_{\mathrm{dust}}$).

To confirm that our results are not sensitive to the chosen SED fitting codes, we also run the analyses using MAGPHYS, following the same procedures as those described in Dudzevičiūtė et al. (2020). The results of MAGPHYS are compared with those of X-CIGALE. To make a more general assessment, we also include fainter SMGs presented in Birkin et al. (2021), where spectroscopic redshifts and multiwavelength data are also available. We find that overall, the physical parameters are comparable between the two codes, except that the X-CIGALE estimated $M_{\mathrm{dust}}$ is, on average, higher by a factor of 1.5 and 1.3 than that in MAGPHYS for our AS2COSPEC sample and the ALESS/AS2UDS sample from Birkin et al. (2021). The 100 Myr averaged SFRs from X-CIGALE are better matched to those from MAGPHYS. However, the 10 Myr averaged SFRs from X-CIGALE are better aligned to the $L_{\mathrm{IR}}$-based SFRs.

We take proper consideration of these differences when making comparisons with other works that employed MAGPHYS (see Appendix A for more details regarding the comparisons). For discussion, we mainly focus on the results based on X-CIGALE, where the inclusion of the AGN component that allows modeling the X-ray photometry is useful in determining the physical properties of our more luminous SMGs.

### 3.2. MBB SED Fitting

In addition to fitting the X-ray–to–radio SED with multiple degree-of-freedom templates, we also employed a single-temperature MBB model to fit the FIR photometry. The MBB fitting method offers the advantage of simplicity and facilitates direct comparisons with results from other samples, in contrast to SED fitting with complex templates.

For the MBB fitting, we utilize photometry from PACS at 100 and 160 μm; SPIRE at 250, 350, and 500 μm; and ALMA at 870, 1250, and 3000 μm. It is worth noting that because of our focus on bright SMGs, more than 80% (15 out of a total of 18) of our sources have at least six FIR photometric data points. Together with spectroscopic redshift measurements, this allows us to obtain constraints that are much tighter than literature samples with only photometric redshifts.

The MBB model is described in the following. The flux density emitted by the dust with a characteristic temperature $T_{\mathrm{dust}}$ and taking the radiative transfer mechanism into account is described by

$$S_\nu = [1 - \exp(-\tau_\nu)][B_\nu(T_{\mathrm{dust}}) - B_\nu(T_{\mathrm{CMB}})]\Delta\Omega(1 + z). \quad (2)$$

For the optically thin case, the dust emission can be approximated as

$$S_\nu = \tau_\nu[B_\nu(T_{\mathrm{dust}}) - B_\nu(T_{\mathrm{CMB}})]\Delta\Omega(1 + z), \quad (3)$$

where the factor $(1 + z)$ is the $k$-correction term, $\Delta\Omega$ is the solid angle of the source, $B_\nu(T_{\mathrm{dust}})$ is the Planck function from dust, $B_\nu(T_{\mathrm{CMB}})$ is the Planck function from the cosmic

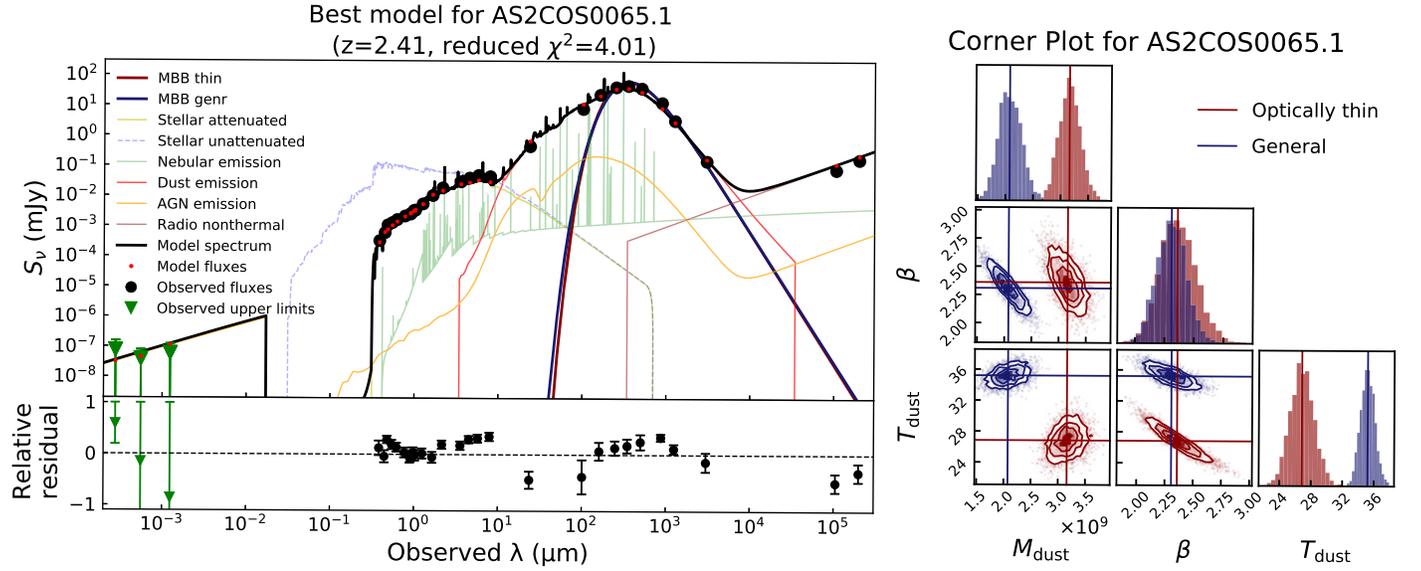

**Figure 2.** Left: an example source of the best-fit SED fitting from X-CIGALE (top panel) and the relative residual of each photometry (bottom panel). Photometry with S/N ⩾ 3 is plotted as the black dots, and that with S/N < 3 is plotted as the green triangles. The black curve is the total best-fit SED model from X-CIGALE, and other colored curves represent the emission from different mechanisms. The dark red and dark blue curves plotted in the FIR regime are, respectively, the best fit of the optically thin and general MBB models (Section 3.2). Right: the corner plot of the MBB fitting from the MCMC of the same source shows the correlation of each parameter in the fitting. Again, red and blue denote the optically thin and general cases, respectively. Each dot represents a set of parameters that construct a model solution, and there are total of 3500 models for each source. The contours are shown at $0.5\sigma$, $1\sigma$, $1.5\sigma$, and $2\sigma$, while the vertical and horizontal lines denote the median value among the total of 3500 sets.

**Table 5**
Physical Parameters from X-CIGALE

| Source | $M_*$ ($\times 10^{10} M_\odot$) | $A_V$ | $L_{IR}$ ($\times 10^{12} L_\odot$) | fracAGN[a] (%) | $SFR_{100Myr}$[b] ($M_\odot\ yr^{-1}$) | $SFR_{IR}$[c] ($M_\odot\ yr^{-1}$) | $M_{dust}$ ($\times 10^9\ M_\odot$) |
|---|---|---|---|---|---|---|---|
| AS2COS0001.1 | 14.7 ± 10.8 | 5.83 ± 0.14 | 35.1 ± 1.8 | 3.7 ± 2.0 | 700 ± 80 | 3630 ± 180 | 2.3 ± 0.1 |
| AS2COS0002.1 | 14.9 ± 2.4 | 4.53 ± 0.3 | 15.3 ± 1.2 | 1.0 ± 0.2 | 590 ± 140 | 1650 ± 130 | 2.8 ± 0.4 |
| AS2COS0006.1 | 8.1 ± 8.7 | 2.92 ± 0.45 | 9.1 ± 0.5 | 1.0 | 440 ± 530 | 980 ± 50 | 5.1 ± 0.3 |
| AS2COS0008.1 | 91.8 ± 8.8 | 3.42 ± 0.04 | 14.5 ± 1.4 | 1.0 ± 0.1 | 1190 ± 290 | 1560 ± 150 | 4.8 ± 0.8 |
| AS2COS0009.1 | 6.1 ± 0.3 | 3.76 ± 0.01 | 6.1 ± 0.3 | 10.0 ± 0.2 | 730 ± 40 | 610 ± 30 | 3.3 ± 0.2 |
| AS2COS0011.1 | 30.3 ± 6.8 | 4.2 ± 0.11 | 20.6 ± 2.5 | 1.3 ± 1.2 | 2670 ± 560 | 2220 ± 270 | 5.2 ± 1.3 |
| AS2COS0013.1 | 14.1 ± 1.0 | 2.04 ± 0.02 | 13.0 ± 0.7 | 1.0 | 350 ± 20 | 1400 ± 70 | 4.0 ± 0.3 |
| AS2COS0014.1 | 40.8 ± 14.9 | 3.76 ± 0.21 | 13.3 ± 2.6 | 1.0 ± 0.3 | 1510 ± 600 | 1430 ± 280 | 4.3 ± 0.5 |
| AS2COS0023.1 | 26.5 ± 10.6 | 3.53 ± 0.34 | 13.8 ± 0.7 | 1.0 | 1690 ± 810 | 1490 ± 70 | 2.8 ± 0.2 |
| AS2COS0028.1 | 3.4 ± 7.8 | 2.53 ± 0.12 | 9.9 ± 1.0 | 1.4 ± 1.2 | 210 ± 20 | 1060 ± 110 | 4.6 ± 1.4 |
| AS2COS0031.1 | 40.8 ± 49.2 | 2.07 ± 0.86 | 9.6 ± 0.5 | 6.3 ± 2.2 | 1110 ± 470 | 1010 ± 50 | 5.8 ± 0.4 |
| AS2COS0037.1 | 40.2 ± 3.5 | 3.54 ± 0.01 | 14.7 ± 0.7 | 1.0 | 2990 ± 150 | 1580 ± 80 | 2.7 ± 0.1 |
| AS2COS0044.1 | 2.8 ± 1.4 | 4.43 ± 0.42 | 10.1 ± 0.6 | 2.3 ± 1.9 | 180 ± 10 | 1090 ± 70 | 5.9 ± 0.3 |
| AS2COS0054.1 | 4.4 ± 1.8 | 2.5 ± 0.1 | 13.0 ± 1.6 | 1.4 ± 1.3 | 300 ± 20 | 1390 ± 170 | 2.7 ± 0.4 |
| AS2COS0065.1 | 9.2 ± 0.5 | 1.9 ± 0.02 | 11.1 ± 0.6 | 1.0 ± 0.1 | 510 ± 30 | 1190 ± 60 | 3.4 ± 0.2 |
| AS2COS0066.1 | 8.4 ± 2.2 | 2.03 ± 0.09 | 7.2 ± 0.8 | 1.2 ± 0.8 | 180 ± 20 | 770 ± 80 | 4.3 ± 0.4 |
| AS2COS0090.1 | 37.6 ± 3.3 | 3.39 ± 0.04 | 4.0 ± 0.3 | 1.0 | 260 ± 80 | 430 ± 40 | 2.7 ± 0.2 |
| AS2COS0139.1 | 13.7 ± 0.7 | 3.91 ± 0.01 | 44.8 ± 2.2 | 1.0 | 1120 ± 60 | 4830 ± 240 | 3.1 ± 0.2 |
| Mean | 22.7 ± 3.1 | 3.35 ± 0.07 | 14.7 ± 0.3 | 2.1 ± 0.2 | 930 ± 80 | 1570 ± 30 | 3.9 ± 0.1 |
| Median | 14.4 ± 6.1 | 3.48 ± 0.32 | 13.0 ± 1.4 | 1.0 ± 0.2 | 640 ± 220 | 1390 ± 150 | 3.7 ± 0.5 |

**Notes.** The uncertainty of the mean is determined by employing the error propagation methodology, which considers the uncertainty associated with each source, and the uncertainty of the median is estimated by the bootstrap method.

[a] Uncertainties less than 0.05 are omitted.
[b] SFR averaged over 100 Myr. See Section 3.1.8 for details.
[c] IR-based SFR (Wuyts et al. 2011), which is the same methodology adopted in Tacconi et al. (2020), calculated by the infrared luminosity emitted by dust ($L_{IR,dust}$) from X-CIGALE.

microwave background (CMB), and $\tau_\nu$ is the optical depth, which can be written as

$$\tau_\nu = \kappa_\nu \Sigma_{dust}, \qquad (4)$$

where $\Sigma_{dust}$ is the dust column density. In our fitting, we assume that the dust geometry is distributed in a homogeneous sphere (Inoue et al. 2020); therefore, the dust column density is

**Table 6**
Physical Parameters from MBB Fitting

| Source | $M_{\rm dust,thin}$ ($\times 10^9 M_\odot$) | $\beta_{\rm thin}$ | $T_{\rm dust,thin}$ (K) | $T_{\rm eff,thin}$[a] (K) | $M_{\rm dust,genr}$ ($\times 10^9 M_\odot$) | $\beta_{\rm genr}$ | $T_{\rm dust,genr}$ (K) | $T_{\rm eff,genr}$[a] (K) | $\lambda_{\rm thick}$[b] ($\mu$m) |
|---|---|---|---|---|---|---|---|---|---|
| AS2COS0001.1 | 1.9 ± 0.2 | 1.8 ± 0.1 | 43 ± 3 | 41 ± 2 | 1.5 ± 0.2 | 1.8 ± 0.1 | 49 ± 2 | 41 ± 2 | 65 ± 10 |
| AS2COS0002.1 | 1.4 ± 0.2 | 2.0 ± 0.1 | 43 ± 3 | 43 ± 2 | 1.1 ± 0.1 | 1.9 ± 0.1 | 51 ± 3 | 43 ± 2 | 64 ± 10 |
| AS2COS0006.1 | 1.8 ± 0.1 | 2.4 ± 0.1 | 32 ± 1 | 35 ± 1 | 1.1 ± 0.1 | 2.3 ± 0.1 | 41 ± 1 | 33 ± 1 | 100 ± 9 |
| AS2COS0008.1 | 7.2 ± 0.8 | 2.3 ± 0.1 | 24 ± 1 | 26 ± 1 | 3.9 ± 0.4 | 2.2 ± 0.1 | 34 ± 1 | 25 ± 1 | 152 ± 13 |
| AS2COS0009.1 | 6.4 ± 0.5 | 1.8 ± 0.1 | 27 ± 1 | 26 ± 1 | 4.9 ± 0.5 | 1.8 ± 0.1 | 33 ± 1 | 26 ± 1 | 118 ± 14 |
| AS2COS0011.1 | 4.2 ± 0.3 | 1.8 ± 0.1 | 37 ± 1 | 36 ± 1 | 3.1 ± 0.2 | 1.8 ± 0.1 | 45 ± 1 | 35 ± 1 | 87 ± 8 |
| AS2COS0013.1 | 2.5 ± 0.1 | 3.2 ± 0.1 | 21 ± 1 | 27 ± 1 | 1.2 ± 0.1 | 3.0 ± 0.1 | 34 ± 1 | 25 ± 1 | 159 ± 9 |
| AS2COS0014.1 | 3.0 ± 0.2 | 2.6 ± 0.1 | 26 ± 1 | 28 ± 1 | 1.8 ± 0.2 | 2.4 ± 0.1 | 36 ± 1 | 28 ± 1 | 131 ± 12 |
| AS2COS0023.1 | 1.6 ± 0.1 | 2.5 ± 0.1 | 32 ± 1 | 35 ± 1 | 1.0 ± 0.1 | 2.4 ± 0.1 | 42 ± 1 | 34 ± 1 | 102 ± 10 |
| AS2COS0028.1 | 3.0 ± 0.3 | 2.1 ± 0.1 | 32 ± 2 | 32 ± 2 | 2.2 ± 0.2 | 2.1 ± 0.1 | 39 ± 2 | 31 ± 2 | 105 ± 13 |
| AS2COS0031.1 | 5.6 ± 0.5 | 2.1 ± 0.1 | 28 ± 1 | 29 ± 1 | 3.4 ± 0.3 | 2.1 ± 0.1 | 37 ± 1 | 28 ± 1 | 131 ± 10 |
| AS2COS0037.1 | 4.2 ± 0.2 | 1.5 ± 0.1 | 40 ± 1 | 35 ± 1 | 3.7 ± 0.2 | 1.4 ± 0.1 | 44 ± 1 | 35 ± 1 | 59 ± 7 |
| AS2COS0044.1 | 8.4 ± 1.5 | 1.5 ± 0.1 | 29 ± 3 | 26 ± 3 | 6.8 ± 1.0 | 1.5 ± 0.1 | 34 ± 3 | 26 ± 4 | 106 ± 16 |
| AS2COS0054.1 | 3.4 ± 0.4 | 2.1 ± 0.1 | 29 ± 1 | 29 ± 1 | 2.4 ± 0.3 | 2.1 ± 0.1 | 36 ± 1 | 29 ± 1 | 108 ± 13 |
| AS2COS0065.1 | 3.2 ± 0.2 | 2.4 ± 0.1 | 27 ± 1 | 29 ± 1 | 2.1 ± 0.2 | 2.3 ± 0.1 | 35 ± 1 | 27 ± 1 | 126 ± 11 |
| AS2COS0066.1 | 4.7 ± 0.5 | 2.3 ± 0.1 | 25 ± 1 | 26 ± 1 | 3.0 ± 0.4 | 2.2 ± 0.1 | 32 ± 1 | 26 ± 1 | 135 ± 13 |
| AS2COS0090.1 | 5.2 ± 0.7 | 2.6 ± 0.1 | 21 ± 1 | 24 ± 1 | 2.6 ± 0.4 | 2.5 ± 0.1 | 30 ± 1 | 23 ± 1 | 157 ± 15 |
| AS2COS0139.1 | 2.5 ± 0.3 | 1.9 ± 0.1 | 37 ± 2 | 37 ± 1 | 2.0 ± 0.3 | 1.9 ± 0.1 | 45 ± 1 | 37 ± 1 | 79 ± 13 |
| Mean | 3.9 ± 0.1 | 2.2 ± 0.03 | 31 ± 0.4 | 31 ± 0.3 | 2.6 ± 0.1 | 2.1 ± 0.02 | 39 ± 0.4 | 31 ± 0.3 | 110 ± 3 |
| Median | 3.3 ± 0.6 | 2.1 ± 0.1 | 29 ± 2 | 29 ± 2 | 2.3 ± 0.4 | 2.1 ± 0.1 | 37 ± 2 | 29 ± 2 | 107 ± 9 |

**Notes.** The uncertainty of the mean is determined by employing the error propagation methodology, which considers the uncertainty associated with each source, and the uncertainty of the median is estimated by the bootstrap method.

[a] The uncertainty of each source is the standard deviation of total of 3500 models from the MCMC method.

[b] Rest-frame wavelength at which the opacity $\tau_{\lambda_{\rm thick}} = 1$.

$\Sigma_{\rm dust} = M_{\rm dust}/(4/3\pi R^2)$, where $M_{\rm dust}$ is the dust mass, which is

$$M_{\rm dust} = \frac{4}{3} \times \frac{S_\nu D_{\rm L}^2}{\kappa_\nu [B_\nu(T_{\rm dust}) - B_\nu(T_{\rm CMB})](1+z)} \quad (5)$$

for the optically thin case.

Different assumptions of the dust distribution geometry, such as a foreground circular obscurer ($\pi R^2$), a spherical shell ($4\pi R^2$), or a homogeneous sphere ($3/4 \times \pi R^2$), would lead to a factor of about 4 difference in dust mass estimates (Inoue et al. 2020). We adopt a homogeneous sphere dust distribution because it yields the closest $M_{\rm dust}$ values to those obtained from X-CIGALE. $\kappa_\nu$ is the frequency-dependent dust opacity, which can be written as

$$\kappa_\nu = \kappa_0(\nu/\nu_0)^\beta, \quad (6)$$

where $\beta$ is the emissivity index and $\kappa_0$ is the emissivity of dust grains per unit mass at a reference frequency $\nu_0$. In this study, we adopt $\kappa_0 = 0.4699$ m$^2$ kg$^{-1}$ at $\nu_0 = 353$ GHz, which is consistent with the value reported by Draine & Li (2007), to ensure a fair comparison with the $M_{\rm dust}$ estimated from X-CIGALE.

In our analyses, we considered both the general and optically thin cases for fitting, with $M_{\rm dust}$, $\beta$, and $T_{\rm dust}$ as the free parameters. For the general case, we also estimated an additional parameter, $\lambda_{\rm thick}$, which is given by

$$\lambda_{\rm thick} = \lambda_0 \left( \kappa_0 \frac{M_{\rm dust}}{4/3\pi R^2} \right)^{1/\beta}. \quad (7)$$

This quantity represents the wavelength at which the opacity $\tau_{\lambda_{\rm thick}} = 1$ (da Cunha et al. 2021) and provides a measure of the optical depth in galaxies, with higher values of $\tau_{\lambda_{\rm thick}}$ indicating a more optically thick scenario.

To quantify the uncertainties and correlations among the parameters, we employ the Markov Chain Monte Carlo (MCMC) method to obtain probability distributions and parameter-space correlations through sampling. We present the best-fit results and corresponding uncertainties from both the optically thin and general model fitting in Table 6. We show an example result in the left panel of Figure 2, where the red and blue curves represent the best-fit MBB models for AS2COS0065.1 for the optically thin and general cases, respectively. Although the shape of the dust emission models appears similar, the best-fit parameters differ; the optically thin model exhibits higher values for $M_{\rm dust}$, while the general model has a higher characteristic dust temperature ($T_{\rm dust}$). The fitting results for all sources are given in Appendix B.

For the optically thin case, the best-fit $\beta$ ranges from approximately 1.5 to 3.2, with a median value and bootstrap uncertainty of $2.1 \pm 0.1$. The dust temperature spans a range from 21 to 43 K, with a median dust temperature of $29 \pm 2$ K. We also estimate the effective dust temperature ($T_{\rm eff}$), obtained from Wien's displacement law for the best-fit MBB model, and find the median value that is the same as the characteristic temperature. The dust masses of our sample SMGs are all on the order of $10^9 M_\odot$, with a median value of $(3.3 \pm 0.6) \times 10^9 M_\odot$.

On the other hand, the dust in the general model is, on average, about 8 K hotter compared to the optically thin model, as the optically thin model assumes that continuum emission escapes from the core, while the general model assumes that only the optically thin surface contributes wavelengths shorter than $\lambda_{\rm thick}$. However, the effective dust temperature is almost the same as that of the optically thin model, which is reasonable considering the similarity in the shape of the best-fit optically thin and general MBB models. The median value of the best-fit dust emissivity index in the general MBB model is $2.1 \pm 0.1$, consistent with that deduced from the optically thin model. The median dust mass from the general MBB model is about $30\% \pm 20\%$ lower but still comparable to the optically thin results when accounting for uncertainties. Again, since there is no significant difference between the best-fit optically thin and general models, the infrared luminosity integrated from the general model is comparable to the value of the optically thin model. Finally, by adopting a typical physical size of $R = 2$ kpc (Hodge et al. 2016), the best-fit $\lambda_{thick}$ ranges from 64 to 159 $\mu$m, with a median value and bootstrap uncertainty of $107 \pm 9$ $\mu$m, consistent with previous estimates for other SMGs (Riechers et al. 2013; Simpson et al. 2017).

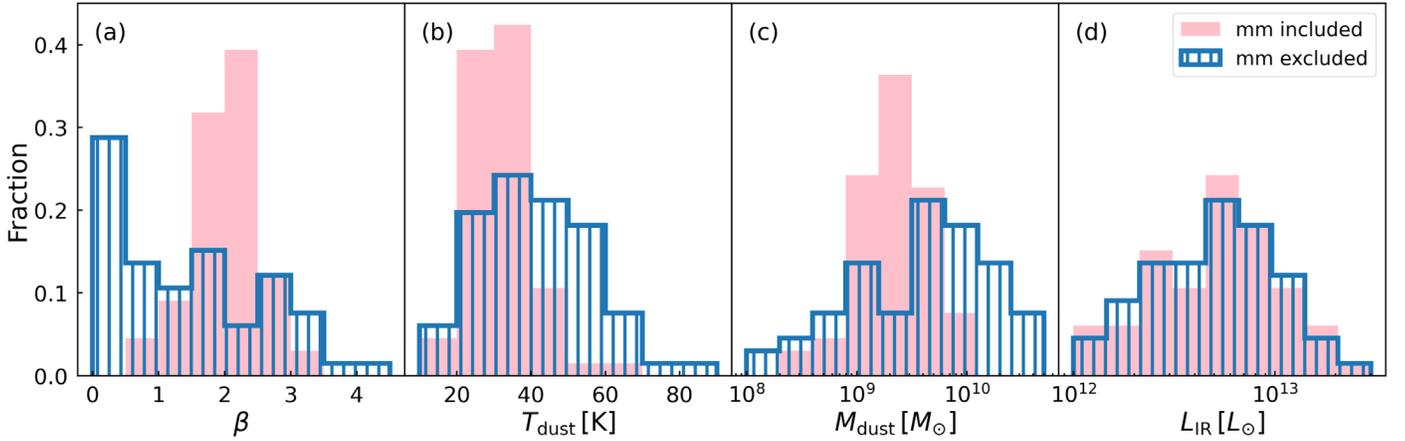

**Figure 3.** Distribution of the best-fit $\beta$, $T_{dust}$, $M_{dust}$, and $L_{IR}$ (from left to right) for the combination of our AS2COSPEC SMGs, as well as the ALESS and AS2UDS sources in Birkin et al. (2021). The pink and blue histograms represent the distributions from millimeter photometry–included and –excluded optically thin MBB fitting. The estimation of $L_{IR}$ does not significantly differ between these two sets. However, the other three dust parameters are more tightly constrained if the millimeter data are included.

In our analysis, we conducted an exercise to gradually exclude millimeter photometry from the fitting process. We discovered that the best-fit parameters are highly sensitive to these longer-wavelength data, which reside on the Rayleigh–Jeans side of the blackbody radiation spectrum. For instance, when the millimeter photometry is omitted, the model constraints are weakened, resulting in lower values of $\beta$ and higher values of $T_{dust}$. In contrast, including the millimeter photometry leads to more tightly constrained best-fit parameters, with a median $\beta_{thin} = 2.1 \pm 0.1$, in close agreement with recent studies (da Cunha et al. 2021). We also applied this exercise to the ALESS and AS2UDS SMGs of Birkin et al. (2021) and found consistent results with our AS2COSMOS SMGs. As illustrated in Figure 3, the fitting without millimeter photometry is unable to accurately constrain the model on the Rayleigh–Jeans tail. Therefore, we caution that comparisons of dust parameters obtained from MBB fitting with other literature should be approached with care, as the fitting outcomes are highly sensitive to the degrees of freedom, especially the longer-wavelength data. We note that similar findings have been reported in da Cunha et al. (2021) for their "well-sampled subset" and the "rest of the sample."

### 3.3. Molecular Gas Mass

The fuel of star formation is the cold molecular gas. Our data allow us to estimate molecular gas masses ($M_{gas}$) using two different tracers: CO luminosity and [C I] luminosity. The summary of our results is provided in Table 7. In the following subsections, we describe how these estimates are obtained.

**Table 7**
Gas Mass

| Source | Line | $M_{gas,CO}$ $(\times 10^{11} M_\odot)$ | $M_{gas,[C\,I]}$ $(\times 10^{11} M_\odot)$ |
|---|---|---|---|
| AS2COS0001.1 | CO(5–4) | $0.8 \pm 0.2 \pm 0.6$ | ⋯ |
| AS2COS0002.1 | CO(5–4) | $1.2 \pm 0.3 \pm 0.9$ | ⋯ |
| AS2COS0006.1 | CO(5–4) | $2.7 \pm 0.3 \pm 2.0$ | ⋯ |
| AS2COS0008.1 | CO(1–0) | $1.3 \pm 0.4 \pm 0.9$ | ⋯ |
| AS2COS0009.1 | CO(3–2) | $1.2 \pm 0.2 \pm 1.1$ | ⋯ |
| AS2COS0011.1 | CO(5–4) | $2.1 \pm 0.2 \pm 1.6$ | $2.5 \pm 0.7 \pm 1.8$ |
| AS2COS0013.1 | CO(1–0) | $1.2 \pm 0.2 \pm 0.9$ | ⋯ |
| AS2COS0014.1 | CO(3–2) | $1.3 \pm 0.2 \pm 1.2$ | ⋯ |
| AS2COS0023.1 | CO(1–0) | $0.8 \pm 0.3 \pm 0.6$ | ⋯ |
| AS2COS0028.1 | CO(3–2) | $0.7 \pm 0.1 \pm 0.6$ | ⋯ |
| AS2COS0031.1 | CO(1–0) | $1.4 \pm 0.4 \pm 1.0$ | $1.5 \pm 0.3 \pm 1.1$ |
| AS2COS0037.1 | ⋯ | ⋯ | ⋯ |
| AS2COS0044.1 | CO(3–2) | $0.7 \pm 0.1 \pm 0.7$ | ⋯ |
| AS2COS0054.1 | CO(1–0) | $1.1 \pm 0.4 \pm 0.8$ | ⋯ |
| AS2COS0065.1 | CO(3–2) | $1.2 \pm 0.1 \pm 1.1$ | ⋯ |
| AS2COS0066.1 | CO(4–3) | $0.8 \pm 0.2 \pm 0.8$ | ⋯ |
| AS2COS0090.1 | CO(4–3) | $0.7 \pm 0.1 \pm 0.7$ | ⋯ |
| AS2COS0139.1 | CO(4–3) | $1.4 \pm 0.1 \pm 1.4$ | ⋯ |
| Mean | ⋯ | $1.2 \pm 0.1 \pm 0.9$ | $2.0 \pm 0.4 \pm 1.5$ |
| Median | ⋯ | $1.2 \pm 0.1$ | $2.0$ |

**Note.** $M_{gas,CO}$ and $M_{gas,[C\,I]}$ represent the gas mass estimated from the CO and [C I] line luminosity, respectively. The first uncertainty represents the measurement uncertainty from line luminosity, while the second uncertainty is the systematic uncertainty from $\alpha_{CO}$ and $r_{J1}$. The uncertainty of the $M_{gas,[C\,I]}$ median is left blank because there are few data to be calculated. For AS2COS0028.1, which has double line detections in our ALMA band 3 survey, we adopt the CO(3–2) measurement for $M_{gas}$ estimation, since the uncertainty of the CO(3–2) measurement is better than that of CO(4–3). For those having CO(1–0) observations, we adopt the $L'_{CO(1-0)}$ measurements from Frias Castillo et al. (2023) to estimate their $M_{gas}$.

#### 3.3.1. CO–$H_2$ Conversion

It is common to estimate molecular gas masses via CO(1–0) luminosities, which can be obtained following

$$M_{gas,CO} = 1.36 \; \alpha_{CO} \; L'_{CO(1-0)}, \quad (8)$$

where $\alpha_{CO}$ is the CO-to-$H_2$ conversion factor in units of $M_\odot$ (K km s$^{-1}$ pc$^2$)$^{-1}$ and the factor of 1.36 accounts for the helium mass.

Five of our SMGs (AS2COS0008.1, AS2COS0013.1, AS2COS0023.1, AS2COS0031.1, and AS2COS0054.1) have CO(1−0) detections reported in the literature; we therefore adopt the $L'_{CO(1-0)}$ measurements from Frias Castillo et al. (2023) for these sources. For other sources that do not have the CO(1−0) measurements, we convert the higher-$J$ transition CO luminosity (Chen et al. 2022b) to $L'_{CO(1-0)}$ using luminosity ratios ($r_{J1}$). We adopt $r_{41}$ and $r_{31}$ from Frias Castillo et al. (2023), where their sample SMGs have similar 870 $\mu$m flux as ours, and $r_{51}$ from Birkin et al. (2021).

For $\alpha_{CO}$, we adopt $\alpha_{CO} = 0.8 \pm 0.6$, which is obtained from the dynamical measurements with a 15% dark matter fraction assumption (Rivera et al. 2018). Overall, 17 of our 18 sample SMGs can have their molecular gas masses estimated via CO luminosity. The $M_{gas,CO}$ measurements are given in Table 7, and we find a median value of $M_{gas,CO} = (1.2 \pm 0.1) \times 10^{11} M_\odot$.

### 3.3.2. [C I]–$H_2$ Conversion

In addition to the CO lines, we also detect two [C I](1−0) lines in two SMGs, AS2COS0011.1 and AS2COS0031.1, and this could also be used as a gas mass tracer. We estimate the gas mass using

$$M_{gas,[C\,I]} = 1.36\, \alpha_{[C\,I]}\, L'_{[C\,I]}, \tag{9}$$

where $L'_{[C\,I]}$ is the [C I] line luminosity in units of K km s$^{-1}$ pc$^2$, $\alpha_{[C\,I]}$ is the [C I]-to-$H_2$ conversion factor in units of $M_\odot$ (K km s$^{-1}$ pc$^2$)$^{-1}$, and the factor of 1.36 accounts for the helium contribution. We adopt the $\alpha_{[C\,I]}$ from Birkin et al. (2021), where they find $\alpha_{[C\,I]}/\alpha_{CO} = 4.4 \pm 0.6$ from the linear fitting in $L'_{[C\,I]} - L'_{CO}$ space. The [C I]-derived gas masses are $(2.5 \pm 0.2) \times 10^{11}$ and $(1.5 \pm 1.2) \times 10^{11} M_\odot$ for AS2COS0011.1 and AS2COS0031.1, respectively, which are consistent with the CO-derived gas masses.

## 4. Discussion

Before discussing the physical properties of our 870 $\mu$m bright SMGs, we first select the physical parameters we adopt in the following subsections. For dust masses, we select those estimated by X-CIGALE. For gas masses, we adopt those estimated by CO emission lines. We do not include AS2COS0037.1 for gas mass analyses, since it does not have any CO line detection. For SFRs, we adopt the infrared luminosity inferred values (Wuyts et al. 2011) in order to make a fair comparison to a sample of MS galaxies presented by Tacconi et al. (2020), where the IR-based SFRs are also adopted. In addition, IR-based SFRs are much less model-dependent compared to the SED-inferred SFRs.

### 4.1. CO Line Properties

We first discuss the CO line properties of the sample. In Figure 4, we plot lines luminosities converted to CO(1−0) using ratios described in Section 3.3.1 versus redshifts, line widths, and infrared luminosities. As a comparison, we also plot similar measurements of fainter SMGs reported in the literature (Bothwell et al. 2013; Birkin et al. 2021), as well as those of a local ULIRG sample (Chung et al. 2009).

First of all, since our AS2COSPEC SMGs are selected to be the brightest in their 870 $\mu$m flux densities, it is not surprising that they are among the most luminous in CO and infrared luminosites. Second, the median FWHM line width is

610 $\pm$ 50 km s$^{-1}$, slightly higher than the measurements of the fainter SMGs, which are reported to be 540 $\pm$ 40 and 500 $\pm$ 60 km s$^{-1}$ by Birkin et al. (2021) and Bothwell et al. (2013), respectively. Finally, we observe that the CO luminosity of our sample does not strongly correlate with any of the plotted properties, where we fit the three properties versus CO line luminosity and find that the best-fit slopes are consistent with zero. This is understandable, since both theory and observations have shown that 870 $\mu$m flux density is a good tracer of dust mass (Dudzevičiūtė et al. 2020; Cochrane et al. 2023), so our sample selection of a pure 870 $\mu$m flux cut effectively means a selection of dust mass and therefore likely a selection of molecular gas mass and thus CO luminosity. The narrow dynamical range in flux density of our sample could also contribute to the flat trend. Indeed, by considering our SMGs and those in the literature together, we find positive correlations between the CO luminosity and the properties plotted in Figure 4.

### 4.2. AGN Properties

Since SMGs are proposed to be the progenitors of local massive quiescent galaxies and coevolve with AGN or quasars at cosmic noon, it is useful to quantify the level at which the SMG and AGN populations overlap. Here, we look at the relationship between AGN and the AS2COSPEC SMGs from the perspectives of SED fitting, X-ray detection, and radio excess.

First, while the fraction of AGN contribution to the total IR luminosity (fracAGN) is allowed up to 50% in our X-CIGALE SED fitting, all of the sources in our sample have a best-fit fracAGN of ∼0%–10%. Only two sources have Bayesian values higher than 5%, while all others are consistent with zero. Overall, the median Bayesian value of fracAGN = $(1.0 \pm 0.2)\%$, suggesting that the infrared luminosity is dominated by the dust emission heated by the stellar emission.

Second, as mentioned in Section 2.2, there are four counterparts found in the Chandra COSMOS-Legacy survey (Civano et al. 2016), which indicates that the X-ray-detected fraction is $22^{+13}_{-18}\%$. To investigate whether the X-ray is mainly powered by AGN or star formation in these sources, we deduce the absorption-corrected rest-frame 0.5–8 keV luminosity ($L_{0.5-8\,keV,corr}$), following the method described in Marchesi et al. (2016), using our CO-based spectroscopic redshift and assuming a power-law energy distribution. The $L_{0.5-8\,keV,corr}$ of these four counterparts spans a range of $(0.3-1.4) \times 10^{44}$ erg s$^{-1}$. As shown in Figure 5(a), the median X-ray–to–IR luminosity ratio $L_{0.5-8\,keV,corr}/L_{IR} = (1.7 \pm 0.8) \times 10^{-3}$ is consistent with the AGN-classified SMG in Alexander et al. (2005). These suggest that the X-ray emission of these sources is mainly powered by AGN rather than star formation.

By taking into account the difference in the X-ray sensitivity, the X-ray-detected fraction appears marginally higher in our sample compared to the less luminous SMG samples such as ALESS and AS2UDS (Wang et al. 2013; Stach et al. 2019). This correlation between IR luminosity and X-ray detection fraction is also reported in the literature using other FIR-selected samples (Kartaltepe et al. 2010; Juneau et al. 2013).

Specifically, in Figure 5(b), we plot the fraction of X-ray counterpart identification rate against infrared luminosity, including our sample, fainter ALESS and AS2UDS SMG samples (Wang et al. 2013; Stach et al. 2019), and the 70 $\mu$m–selected galaxies (Kartaltepe et al. 2010). We also plot the correlation curve presented by Kartaltepe et al. (2010), where they cross-match their sources to both the XMM and Chandra surveys. To make a fairer comparison with the SMG samples, we reconstruct the correlation curve of Kartaltepe et al. (2010) by first cutting the sample to only include $z > 1$ sources and then matching the X-ray depth to the Chandra COSMOS-Legacy survey (Civano et al. 2016). The matching of the X-ray depth is also applied to the ALESS and AS2UDS samples to assess the X-ray detection fractions. The reconstructed correlation curve of the 70 μm–selected sources is broadly consistent with the results based on SMGs.

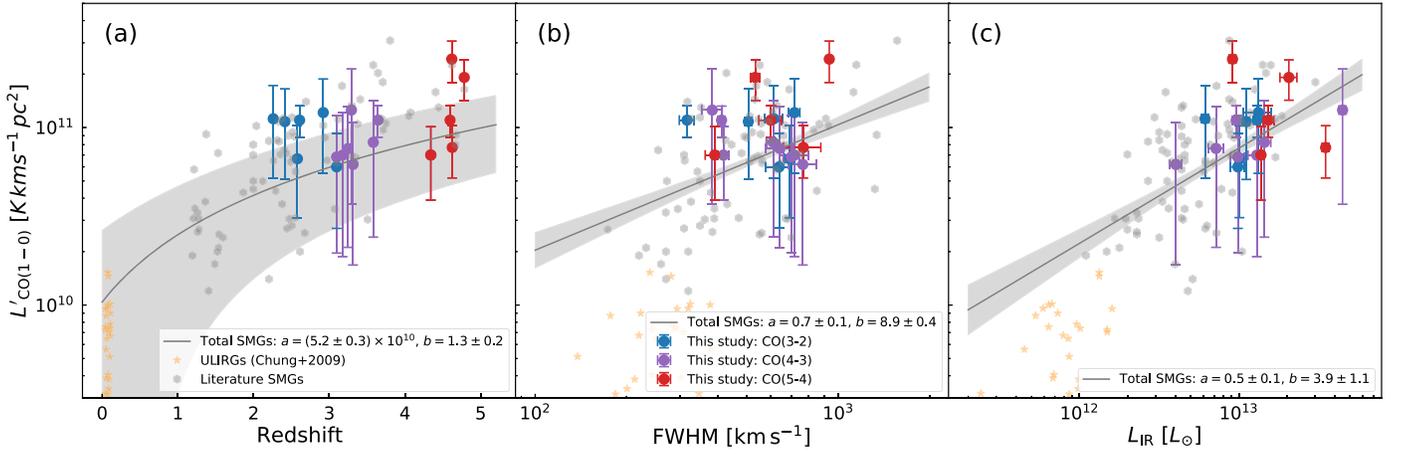

**Figure 4.** (a) $L'_{CO(1-0)}$ vs. redshift of our $S_{870}$-bright SMGs, the literature SMGs (Bothwell et al. 2013; Birkin et al. 2021), and the local ULIRGs (Chung et al. 2009). We fit the literature SMGs by the model in the form of $L'_{CO(1-0)} = a \times [(1+z)/(1+z_{med})]^b$, and the best-fit curve is shown as the gray line. (b) $L'_{CO(1-0)}$ vs. the FWHM of CO emission lines for our sample, as well as the literature data. The literature SMGs are fitted by a linear relation with the form $\log L'_{CO(1-0)} = a \times \log FWHM + b$, and the best fit is shown as the gray line. (c) $L'_{CO(1-0)}$ vs. $L_{IR}$ of our sample and the literature data. Again, we fit the literature SMGs by a linear model $\log L'_{CO(1-0)} = a \times \log L_{IR} + b$ and present the best fit with a gray line. For all panels, the blue, purple, and red dots represent our AS2COSPEC SMGs with different CO transitions, the gray hexagons are the SMGs from Birkin et al. (2021) and Bothwell et al. (2013), the orange stars are the local ULIRGs from Chung et al. (2009), the gray lines are the best-fit functions including all SMG samples, and the gray bands denote the 68% confidence interval of the fitting.

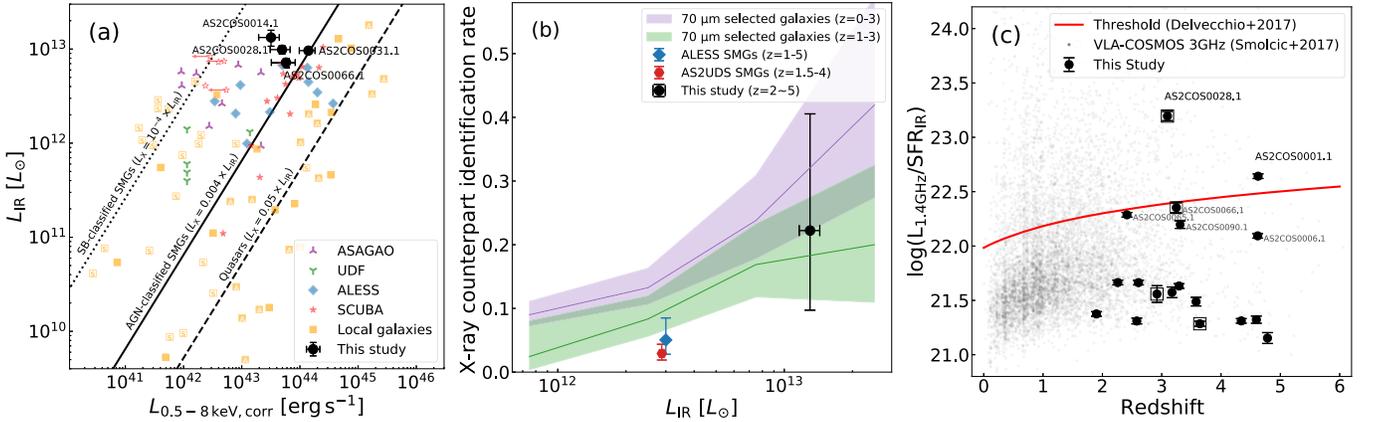

**Figure 5.** (a) Absorption-corrected rest-frame X-ray luminosity vs. infrared luminosity. Our four sources that have X-ray counterparts are black dots, the ASAGAO millimeter galaxies (Ueda et al. 2018) are purple upward triangles, the UDF sources (Dunlop et al. 2016) are green downward triangles, the ALESS SMGs from Wang et al. (2013) are blue diamonds, the SCUBA SMGs (Alexander et al. 2005) are red stars (SB-classified sources are open and AGN-classified sources are filled), and local galaxies are yellow squares classified as AGN-dominated (labeled "A") or star formation-dominated (labeled "S") by Rigopoulou et al. (1999) and Tran et al. (2001). The dotted line shows the mean relation of the SB-classified SMGs from Alexander et al. (2005), the solid line represents the median relation of the AGN-classified SMGs from Alexander et al. (2005), and the dashed line demonstrates the median relation of quasars studied by Elvis et al. (1994). (b) X-ray identification rate as a function of infrared luminosity. Our AS2COSPEC sample is the black dot, the ALESS sample from Wang et al. (2013) is the blue diamond, and the AS2UDS sample from Stach et al. (2019) is the red hexagon. The error bars are calculated by Poisson statistics (Gehrels 1986), and the asymmetric error propagation is implemented (Gobat 2022). The purple and green trends are the X-ray-detected fraction of the 70 μm–selected galaxies obtained from Kartaltepe et al. (2010) with different redshift cuts. (c) Rest-frame 1.4 GHz luminosity to IR-based SFR as a function of redshift. Our AS2COSPEC SMGs are the black dots (sources with X-ray counterparts are boxed), the VLA-COSMOS 3 GHz sources (Smolčić et al. 2017) are the gray dots, and the 3σ deviation threshold of the 3 GHz sources is shown as the red curve (Delvecchio et al. 2017).

Third, we identify the AGN by checking the excess of radio emission. Specifically, we compute the 1.4 GHz luminosity to the IR-based SFR ratio following the method described in Smolčić et al. (2017). By adopting the redshift-dependent threshold, which is the 3σ deviation to the VLA-COSMOS 3 GHz sources (Delvecchio et al. 2017), only two sources in our sample (AS2COS0001.1 and AS2COS0028.1) are identified as radio-excess sources, as shown in Figure 5(c).

Overall, from the points of view of multiwavelength SED fitting, X-ray counterpart, and excess of radio emission, we find that AGN play a minor role in each perspective. We summarize the identification of AGN in Table 8, where there is no overlap for all sources except AS2COS0028.1 and AS2COS0031.1. However, the low detection rate in the X-ray and the low excess rate in the radio do not preclude the existence of heavily obscured AGN, which may have been missed significantly by X-ray surveys as suggested by recent studies (e.g., Carroll et al. 2023). More mid-infrared data such as those from the James Webb Space Telescope (JWST) would help improve our understanding with regard to the relationship between AGN and SMGs.

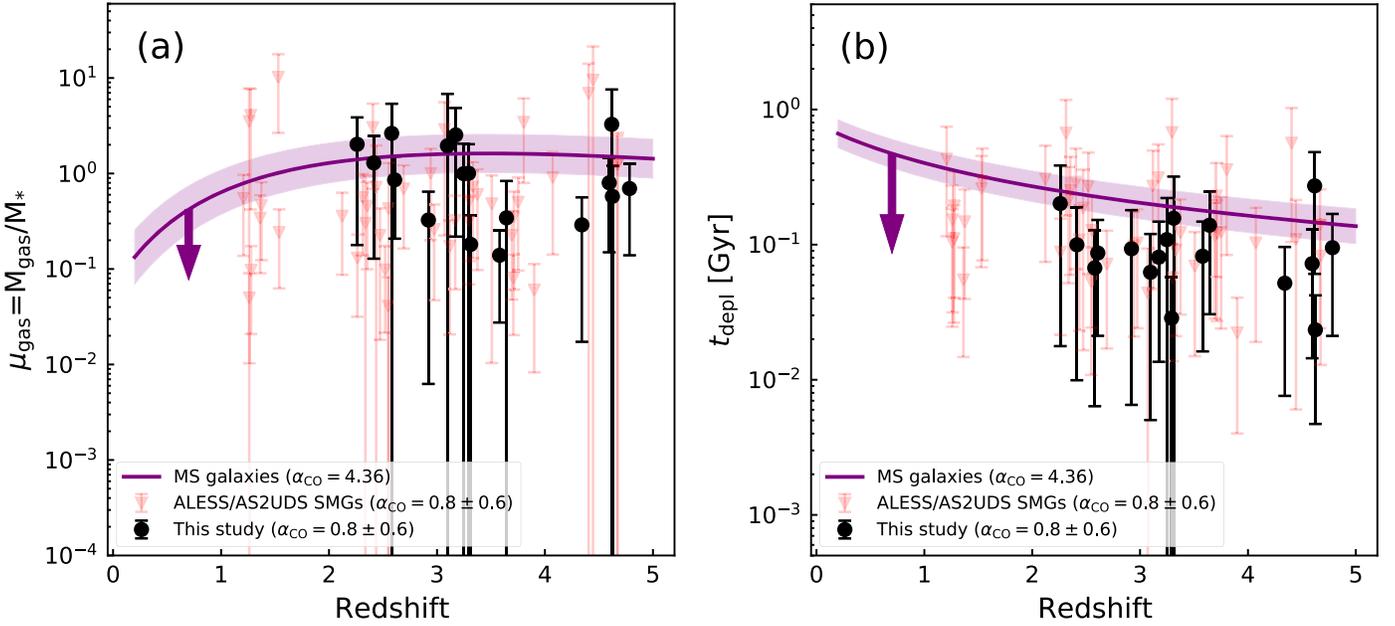

**Figure 6.** Scaling relations of the gas fraction ($\mu_{gas} = M_{gas}/M_*$) with redshift (a) and the gas depletion timescale ($t_{depl} = M_{gas}/SFR$) with redshift (b). In both plots, black dots are our AS2COSPEC SMGs, red triangles are the ALESS/AS2UDS SMGs (Birkin et al. 2021), and purple lines represent the predictions from the best-fit results of the MS galaxies (Tacconi et al. 2020) by adopting the median $\delta$MS and $M_*$ of our sample. The purple downward arrows demonstrate the systematic shift of the MS relations when adopting $\alpha_{CO} = 0.8$.

### 4.3. Gas Fraction and Depletion Time

To quantify gas properties, we derive two parameters: the molecular gas fraction, defined as

$$\mu_{gas} = \frac{M_{gas}}{M_*}, \qquad (10)$$

which describes the size of the molecular gas reservoir normalized to the stellar mass, and the gas depletion timescale, defined as

$$t_{depl} = \frac{M_{gas}}{SFR}, \qquad (11)$$

which is the timescale for a galaxy to convert its molecular gas to stars through star formation, assuming no gas replenishment.

Overall, for the AS2COSPEC SMGs, we find a range of gas fraction from 0.1 to 3.3, with a median of $\mu_{gas} = 0.9 \pm 0.3$, indicating that the molecular gas mass is, on average, comparable to the stellar mass. In contrast, the fainter SMGs in the same redshift range as ours from Birkin et al. (2021) have a median $\mu_{gas} = 0.5 \pm 0.1$, which is about $2\sigma$ lower. This is a combined effect of fainter SMGs being, on average, ~20% more massive in stellar mass and ~20% less massive in molecular gas mass. On the other hand, the range for the depletion time for the AS2COSPEC SMGs spans from 20 to

**Table 8**
**AGN Identification**

| Source | fracAGN ⩾ 5% | X-Ray | Radio Excess |
|---|---|---|---|
| AS2COS0001.1 | ... | ... | ✓ |
| AS2COS0002.1 | ... | ... | ... |
| AS2COS0006.1 | ... | ... | ... |
| AS2COS0008.1 | ... | ... | ... |
| AS2COS0009.1 | ✓ | ... | ... |
| AS2COS0011.1 | ... | ... | ... |
| AS2COS0013.1 | ... | ... | ... |
| AS2COS0014.1 | ... | ✓ | ... |
| AS2COS0023.1 | ... | ... | ... |
| AS2COS0028.1 | ... | ✓ | ✓ |
| AS2COS0031.1 | ✓ | ✓ | ... |
| AS2COS0037.1 | ... | ... | ... |
| AS2COS0044.1 | ... | ... | ... |
| AS2COS0054.1 | ... | ... | ... |
| AS2COS0065.1 | ... | ... | ... |
| AS2COS0066.1 | ... | ✓ | ... |
| AS2COS0090.1 | ... | ... | ... |
| AS2COS0139.1 | ... | ... | ... |

**Note.** AGN identification from multiwavelength SED fitting, X-ray counterpart, and excess of radio emission.

270 Myr, with a median $t_{depl}$ of $90 \pm 10$ Myr, which is about 25% lower than the median value of $120 \pm 30$ Myr found for the fainter SMGs from Birkin et al. (2021) at the same redshift range. Our finding that both the increase of gas supply and the increase of star formation efficiency play a role in driving galaxies to move above the MS is consistent with recent population studies of star-forming galaxies at similar epochs (Tacconi et al. 2020; Scoville et al. 2023).

In Figure 6, we plot the evolution of the gas fraction and depletion time deduced by Tacconi et al. (2020), which is based on a sample of 2052 star-forming galaxies with $\delta$MS = log(SFR/SFR$_{MS}$) ranging from $-2.6$ to $+2.2$ at $z = 0$–5.5.

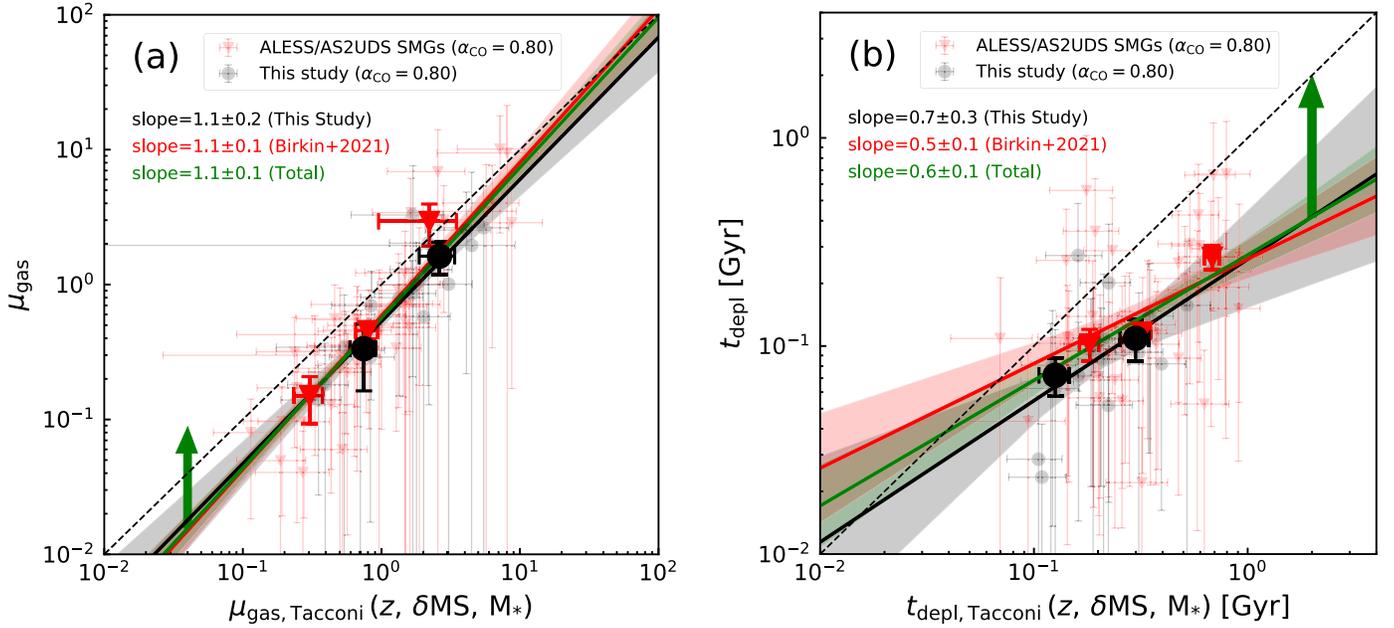

**Figure 7.** The $\mu_{gas}$ and $t_{depl}$ comparisons between the observational measurements (our AS2COSPEC SMGs in black and the ALESS/AS2UDS SMGs from Birkin et al. 2021 in red) and predictions from the MS star-forming galaxies (Tacconi et al. 2020) are plotted in panels (a) and (b). For both plots, individual measurements are plotted as small transparent symbols, while the binned median values are shown as the large symbols. There are two and three bins for our sample and the literature sample, respectively. Uncertainties for the binned data are derived from the bootstrap method. The linear best fits to our sample (black), the literature sample (red), and the combination of both samples (green) are also plotted, with the upward arrows pointing to the systematic offset if adopting $\alpha_{CO} = 4.36$.

They claim that $\mu_{gas}$ and $t_{depl}$ can be effectively described as a function of three parameters, $z$, $\delta MS$, and $M_*$, and the correlations can be generally explained by the Toomre stability criteria (Toomre 1964).

By comparing the functional curves with SMGs, we find that SMGs are broadly consistent with the relationships of the MS galaxies. Upon closer inspection, some data points appear to deviate from the scaling relationship of the MS galaxies. However, it is important to note that since the correlation functions are dependent on three parameters ($z$, $\delta MS$, and $M_*$), the correlation curves in Figure 6 are plotted by assuming median values of the SMG samples for $\delta MS$ and $M_*$. In addition, these scaling relations of the MS are obtained by using $\alpha_{CO} = 4.36$ for estimating the gas mass, which is systematically higher than what we calculated for the SMGs. It is thus nontrivial in this case to compare data points against the relationships using simple statistical tests.

To facilitate a more straightforward comparison, for each source, we use the relationships reported by Tacconi et al. (2020) and compute the expected $\mu_{gas}$ and $t_{depl}$ given its $z$, $\delta MS$, and $M_*$. We plot the results against the measured $\mu_{gas}$ and $t_{depl}$ in Figure 7. In spite of the normalization caused by the $\alpha_{CO}$, we find a good agreement between the expected and measured $\mu_{gas}$, where the best-fit slopes for our AS2COSPEC sample, the ALESS/AS2UDS sample (Birkin et al. 2021), and the combination of both samples are consistent with unity. This infers that the correlation functions for $\mu_{gas}$ proposed by Tacconi et al. (2020) are applicable to the SMGs. As for $t_{depl}$, the best-fit slopes to the three subsets are all sublinear, which suggests that the gas depletion timescale for the SMG population is shorter compared to the one deduced based on the correlation proposed by Tacconi et al. (2020). These results suggest that while on average, SMGs share a common behavior with fainter and less massive star-forming galaxies in terms of their gas fractions, SMGs are more efficient in star formation, possibly due to being preferably located in dense environments where dynamical interactions are more common (Chen et al. 2022a; Rujopakarn et al. 2023; Smail et al. 2023).

We note that if the 100 Myr averaged SFR deduced from X-CIGALE is adopted, the median $t_{depl}$ of our AS2COSPEC SMGs and the fainter SMGs in the same redshift range as ours from Birkin et al. (2021) are $200 \pm 60$ and $270 \pm 80$ Myr, systematically higher than that derived using IR-based SFR. However, our AS2COSPEC sample is still 25% lower than the SMGs from Birkin et al. (2021). As for the $t_{depl}$ comparison to the MS star-forming galaxies from Tacconi et al. (2020), the best-fit slopes of our sample, the sample from Birkin et al. (2021), and the total SMG samples are $0.8 \pm 0.5$, $0.9 \pm 0.2$, and $0.9 \pm 0.2$. While this could suggest that SMGs may have a similar star formation efficiency as the less massive and less active star-forming galaxies, we stress that the adoption of different methods of deducing SFRs could make this conclusion misleading. We therefore make our conclusion based on the IR-based SFRs.

### 4.4. Dust Properties

#### 4.4.1. Dust Fraction

We begin the discussion about dust by deriving the dust fraction ($\mu_{dust} = M_{dust}/M_*$), meaning the dust-to-stellar mass ratio. The dust production mechanisms include ejection from supernovae (SNe) and asymptotic giant branch stars, accretion from the ISM, and infalling from the intergalactic medium, while the mechanisms of dust destruction include SN shock, stellar feedback, and AGN feedback. The observed dust-to-stellar mass ratio is therefore related to the balance between dust production and destruction and could potentially be used for constraining feedback physics in theoretical models.

In our sample, we find a median dust fraction of $\mu_{dust} = (2.1 \pm 1.0) \times 10^{-2}$. This is in contrast to the median dust fraction of $\mu_{dust} = (6.7 \pm 1.4) \times 10^{-3}$ obtained from the ALESS and AS2UDS SMGs of Birkin et al. (2021), where we apply the same analysis techniques to derive the dust properties. Similarly, Donevski et al. (2020) analyzed a sample of 300 Herschel-selected DSFGs up to $z \approx 5$ and found a median dust fraction of $\mu_{dust} = (6.6 \pm 0.5) \times 10^{-3}$ using the same SED package, X-CIGALE, as our analysis. Furthermore, Dudzevičiūtė et al. (2020) determined a median dust fraction of $\mu_{dust} = (8.2 \pm 0.4) \times 10^{-3}$ for 707 AS2UDS SMGs, where the systematic difference between X-CIGALE and MAGPHYS in $M_{dust}$ is corrected. Comparing our sample to the SMG samples in the literature, our dust fraction is approximately four times higher. This difference is likely due to sample selection, as sources with brighter submillimeter flux tend to have higher dust masses.

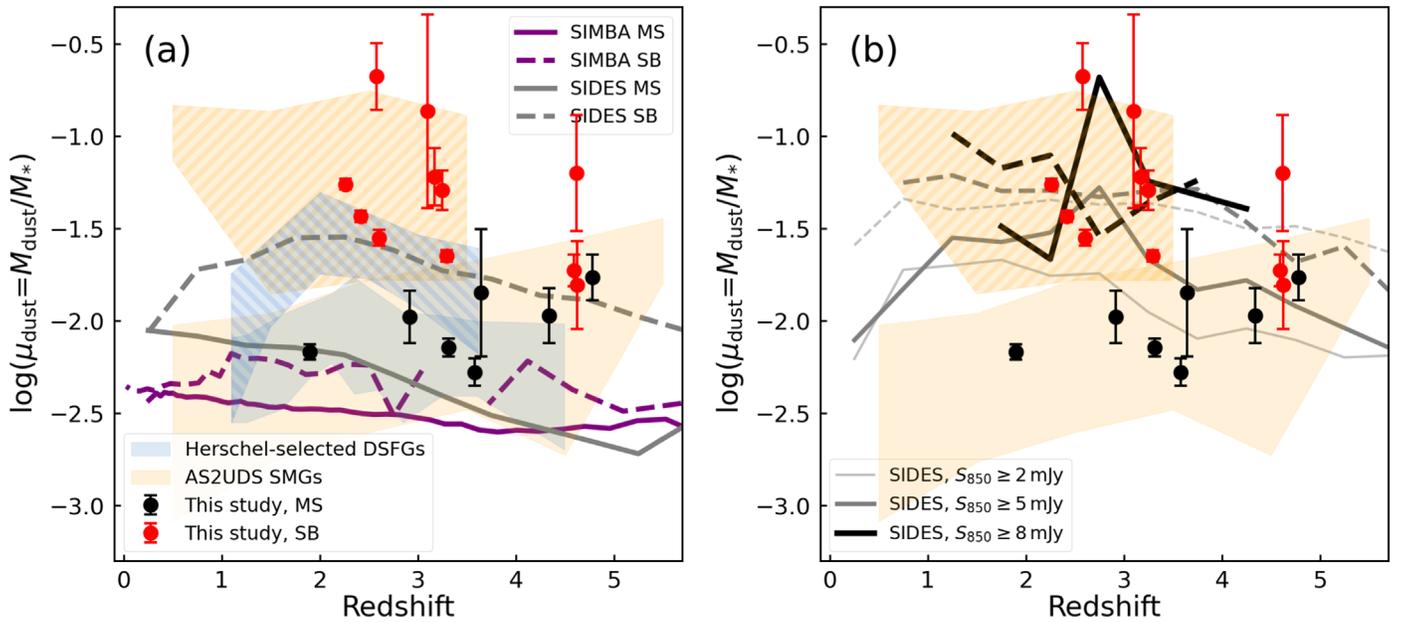

**Figure 8.** (a) Dust fraction ($\mu_{dust} = M_{dust}/M_*$) as a function of redshift. Black dots represent our AS2COSPEC MS SMGs, and red dots are the AS2COSPEC SB SMGs. Distributions of the Herschel-selected DSFGs (Donevski et al. 2020) and the AS2UDS SMGs (Dudzevičiūtė et al. 2021) are plotted as the blue and yellow shaded areas, where the MS subsamples are the nonhatched regions and the SB ones are the hatched ones. To remove the dust mass dependency due to SED fitting codes, systematic correction between MAGPHYS and X-CIGALE is applied to the AS2UDS SMGs. Simulated predictions from SIMBA (Davé et al. 2019) and SIDES (Béthermin et al. 2022) are plotted as the purple and gray lines, where the solid lines represent the MS galaxies and the dashed lines are the SB ones. Same as panel (a), showing the SIDES predictions with different $S_{850}$ selection. Due to the small sample size, we do not plot the evolution of sources brighter than 8 mJy. Curves with a brighter $S_{850}$ cut are consistent with our sources, showing that, compared to literature samples, our high $\mu_{dust}$ measurements could be due to the sample selection.

Figure 8(a) displays $\mu_{dust}$ plotted against redshift for our sample, the AS2UDS SMGs (Dudzevičiūtė et al. 2020), and the Herschel-selected DSFGs (Donevski et al. 2020). Following the method in Donevski et al. (2020) and adopting the MS correlation from Speagle et al. (2014), each sample is split into MS and SB subsamples, where the boundary between two subsets is defined as $\Delta MS (= SFR/SFR_{MS}) = 4$. Overall, we find a good agreement in $\mu_{dust}$ between SMG samples, but it appears higher than the Herschel-selected DSFGs in the SB subset. This could be understood as that the 850 $\mu$m selection tends to preferentially select higher dust mass sources.

To compare with model predictions, we also include two simulated data sets, which are also split into SB and MS subsamples in Figure 8(a). They are the cosmological galaxy formation and evolution model SIMBA (Davé et al. 2019) and the phenomenological model SIDES (Béthermin et al. 2022). While SIMBA provides dust mass, we calculate dust masses for SIDES sources using the model of Draine & Li (2007) and the provided infrared photometry. To ensure a fair comparison between observation and simulation, we only plot the simulated galaxies that have similar physical properties to our sources in Figure 8(a), namely, unlensed sources with log $(M_*/M_\odot) \geqslant 10.44$ and log $(sSFR/yr^{-1}) \geqslant -9.32$. The binned averages of the MS and SB samples from both simulations are shown as the solid and dashed lines in Figure 8.

We find that, for both MS and SB galaxies, the $\mu_{dust}$ predictions from SIMBA are systematically lower than the observed values. The lack of dust-rich galaxies in SIMBA was also presented and discussed in Li et al. (2019). By comparing the observed gas-to-dust ratio and gas fraction with SIMBA predictions, Donevski et al. (2020) suggest that this discrepancy could be due to a timescale for dust growth. The dust growth timescale is inversely proportional to gas-phase metallicity, and most galaxies in SIMBA, especially at $z \gtrsim 2$, have subsolar metallicity (Figure 3 in Li et al. 2019), which suggests inefficiency in dust growth. The reason for SIMBA having too low of a gas-phase metallicity is unclear. On the other hand, the dust growth timescale is also inversely proportional to the cold gas surface density. According to a recent investigation conducted by Cochrane et al. (2023), it is proposed that the intensity of AGN feedback has a notable influence on the FIR sizes of central star-bursting regions, consequently affecting the cold gas surface density. Specifically, stronger AGN feedback is linked to larger sizes and lower cold gas densities. This could suggest that AGN feedback in SIMBA is too strong, which drives too low of a cold gas density, leading to too long of a dust growth timescale. Comparing the observed submillimeter sizes with those predicted by SIMBA would help confirm this hypothesis.

On the other hand, we find that SIDES predictions are in good agreement with Herschel-selected DSFGs, confirming the results reported by Donevski et al. (2020). However, they do not fully agree with SMGs. First, the predicted values are, on average, at the lower end of the measurements for the SB sources. Second, the trend in redshift for the MS sources goes in the opposite direction between the predicted and observed values; as a result, the deviation between the SIDES predictions and the observed values becomes more significant at $z \gtrsim 4$. To test if these discrepancies are mainly caused by the 850 $\mu$m selection, in Figure 8(b), we plot SIDES model curves based on various 850 $\mu$m flux density cuts. We find that indeed, by imposing the 850 $\mu$m flux cuts, the predicted values are in better agreement with the measurements for the SB sources. However, for MS sources, SIDES predicts higher $\mu_{dust}$, up to $\sim 0.5$ dex for the brightest SMGs. By comparing the distributions of dust and stellar mass, we conclude that the overestimating $\mu_{dust}$ is mainly driven by about a factor of 2 lower stellar mass in SIDES SMGs, which could be caused by the large uncertainties of the most massive end of the stellar mass functions. Lastly, the decreasing trend in redshift predicted by SIDES can still be observed in Figure 8(b) despite adopting 850 $\mu$m flux density cuts. This could be due to the fact that SIDES has a limited predicting power at $z \gtrsim 4$, since it relies on observed correlations and stellar mass functions, which are mostly not well constrained at $z \gtrsim 4$.

### 4.4.2. Gas-to-dust Mass Ratio

Our data also allow us to estimate the gas-to-dust ratio, $\delta_{gdr}$, which has been observed to correlate with redshift (Saintonge et al. 2013; Péroux & Howk 2020) and metallicity (Leroy et al. 2011; De Vis et al. 2019). From the Spitzer Infrared Nearby Galaxy Survey, the $\delta_{gdr}$ is roughly 110 (Draine et al. 2007). By adopting $\alpha_{CO} = 1$ and using the MBB FIR fitting method, Swinbank et al. (2013) obtain an average $\delta_{gdr} = 90 \pm 25$ for the ALESS SMGs. If adopting the same dust distribution geometry of our analyses, $\delta_{gdr}$ would become $68 \pm 19$ for the ALESS SMGs in Swinbank et al. (2013).

In our analyses, we estimated $M_{dust}$ using three different methods: X-CIGALE X-ray–to–radio SED fitting (Section 3.1), optically thin fitting, and general MBB fitting to the FIR photometry (Section 3.2). Based on these three dust mass estimations, we calculated the corresponding $\delta_{gdr}$, and the median values are $32 \pm 3$, $44 \pm 10$, and $56 \pm 10$, respectively. After rescaling the $\alpha_{CO}$ to 0.8 and correcting for the ~1.5 times difference in dust mass between X-CIGALE and MAGPHYS (Appendix A), the MAGPHYS-based median, $\delta_{gdr} = 34 \pm 4$, of the ALESS and AS2UDS SMGs (Birkin et al. 2021) is consistent with our X-CIGALE-based median ($\delta_{gdr} = 32 \pm 3$). Our MBB-based median values are also consistent with the estimates of the ALESS and AS2UDS SMGs when the rest-frame 870 $\mu$m luminosity is used as a tracer of $M_{dust}$ (Birkin et al. 2021). The relatively low gas-to-dust ratio could suggest supersolar gas-phase metallicity, which is in line with SMGs being heavily dust-enriched and consistent with some recent results (Birkin et al. 2023; Eales et al. 2023; Peng et al. 2023).

### 4.4.3. Dust Temperature

In Figure 9(a), we present the redshift dependence of $L_{IR}$. We find that, compared to fainter SMGs, our AS2COSPEC SMGs are more luminous in $L_{IR}$ at given redshifts, as expected, and we do not find significant redshift dependency in $L_{IR}$ in our sample, which might be attributed to selection effects.

In Figure 9(b), we plot dust temperature ($T_{dust}$) against $L_{IR}$, and we also include results from fainter SMGs from AS2UDS (Dudzevičiūtė et al. 2020) and Herschel-selected infrared luminous galaxies ($L_{IR} > 10^{10}L_{\odot}$) at $0.1 < z < 2$ (Symeonidis et al. 2013). To ensure proper comparisons to the AS2UDS SMGs, we follow the methodology presented by Dudzevičiūtė et al. (2020) for the MBB fittings, meaning that they are not corrected for the CMB, and a fixed $\beta = 1.8$ and optically thin case are assumed.

Within a given $L_{IR}$ bin, the $T_{dust}$ of our bright SMGs is lower than that of the AS2UDS SMGs. This difference in temperatures likely arises from a selection bias in our sample, as our selection favors sources with higher dust masses due to the positive correlation between $M_{dust}$ and $S_{870}$. Consequently, we preferentially select sources with colder $T_{dust}$ at a fixed $L_{IR}$. To demonstrate this selection effect more clearly, in Figure 9(b), we plot the selection function as the gray hatched area. This selection function is constructed using MBB models that have the same $S_{870}$ range as our sample while fixing the redshift at the median value ($z = 3.3$) and setting $\beta$ to 1.8.

To discuss the correlation between dust temperature and redshift, we focus our analysis on a specific region of the parameter space shown in Figure 9(a), selecting sources within an $L_{IR}$ slice ($L_{IR} = (4–12) \times 10^{12}L_{\odot}$) and a redshift range ($z = 2.0–4.0$), which represents the region where most of our sources are located and exhibits less dependence on both $L_{IR}$ and redshift. This allows us to investigate the evolution of $T_{dust}$ more effectively. In our sample, we find no significant evolution of $T_{dust}$, as illustrated in Figure 9(c). Note that the correlation observed for the AS2UDS sample shown in Figure 9 is likely influenced by the $L_{IR}$–$z$ dependence; therefore, it is not solely driven by the $T_{dust}$–$z$ correlation.

### 4.4.4. Dust Emissivity Index

The dust emissivity index ($\beta$) is reflected in the slope of the MBB model in the Rayleigh–Jeans tail. In the Rayleigh–Jeans regime, the slope is determined by the contribution from the Planck function ($\alpha_{pl}$), given by $S_{\nu} \propto \nu^{\alpha_{pl}}$, and the dust emissivity ($\beta$), given by $S_{\nu} \propto \nu^{\beta}$. A steeper slope corresponds to a higher value of $\beta$. We derive a median $\beta$ of $2.1 \pm 0.1$ for our sample, which is consistent with the value of $2.0 \pm 0.1$ obtained in Birkin et al. (2021). The slightly higher $\beta$ in our sample may be influenced by the selection bias toward lower $T_{dust}$ sources. Despite this, $\beta$ is comparable between our brightest SMGs and the literature SMGs within the uncertainties (Magnelli et al. 2012; Casey et al. 2021; da Cunha et al. 2021).

Figures 9(e) and (f) present the variations of $\beta$ as a function of $L_{IR}$ and redshift for our sample, the literature SMGs (Birkin et al. 2021; Bendo et al. 2023; McKay et al. 2023), and the local Herschel-selected galaxies from Smith et al. (2013). We perform linear fits to the parameters and find no correlations with either $L_{IR}$ or redshift, suggesting that the dust grain properties do not significantly correlate with $L_{IR}$ or redshift in galaxies. We again emphasize the importance of millimeter photometry in constraining the MBB model, as presented in Figure 3.

Several dust grain properties, such as dust grain size and composition, can influence the value of $\beta$. Theoretical studies have demonstrated an inverse relationship between $\beta$ and dust grain size for different types of dust compositions (Ysard et al. 2019). For instance, $\beta$ is approximately 1 for millimeter- to

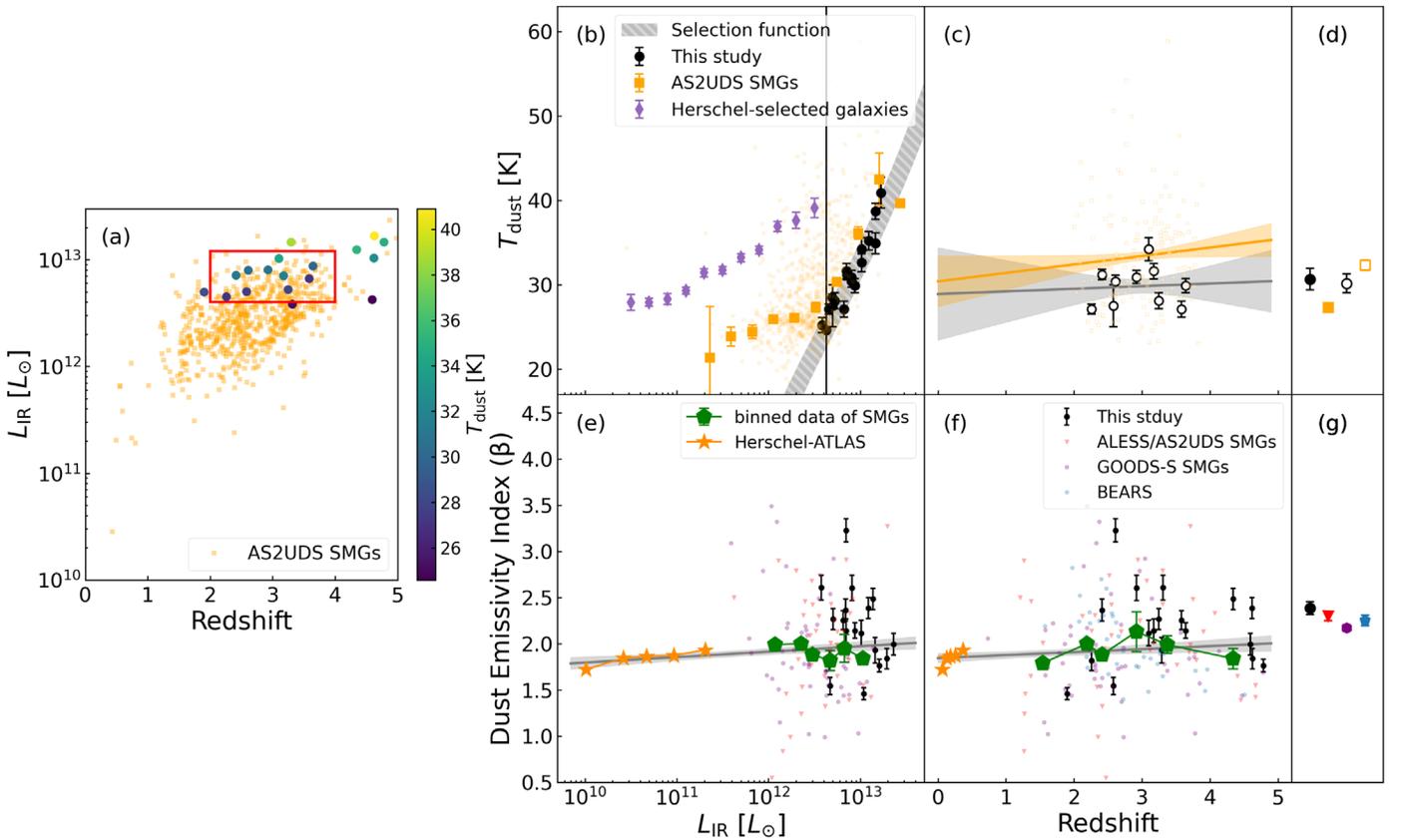

**Figure 9.** (a) $L_{IR}$ as a function of redshift of our sample colored by their $T_{dust}$ and the AS2UDS SMGs (Dudzevičiūtė et al. 2020; orange squares). The red box encloses sources, which are plotted in panel (c), at a given $L_{IR}$ bin. (b) $T_{dust}$ vs. $L_{IR}$ of our sample (black dots), the AS2UDS SMGs (Dudzevičiūtė et al. 2020; small transparent orange squares), and lower-redshift Herschel-selected galaxies (Symeonidis et al. 2013; purple diamonds). Larger yellow squares are the median values of AS2UDS SMGs from Dudzevičiūtė et al. (2020) at several $L_{IR}$ bins. The gray hatched area represents the selection function of our AS2COSPEC SMGs. (c) $T_{dust}$ vs. redshift of sources within the red box in panel (a). No significant evolution is observed in our sample. (d) Median $T_{dust}$ values for different subsets as depicted in panels (b) and (c). For the plots shown in panels (a)–(d) that compare to Dudzevičiūtė et al. (2020), we obtain the $T_{dust}$ and $L_{IR}$ of our sample from the non-CMB optically thin MBB fitting with $\beta$ fixed to 1.8 using 100–870 $\mu$m photometry, consistent with the methodology in Dudzevičiūtė et al. (2020). (e) $\beta$ vs. $L_{IR}$ for our sample (black dots), the ALESS/AS2UDS SMGs (Birkin et al. 2021; red triangles), the GOODS-S SMGs (McKay et al. 2023; purple hexagons), and the binned data of the local Herschel-selected galaxies (Smith et al. 2013; orange stars). A linear fit is applied, revealing no correlation with a slope of $0.06 \pm 0.03$. (f): $\beta$ vs. redshift for the same samples shown in panel (e), and measurements from the lensed Herschel BEARS program (Bendo et al. 2023) are plotted as blue pentagons. Again, a linear fit is performed, indicating no evolution with a slope of $0.03 \pm 0.02$. (g) Median $\beta$ of different subsets in panels (e) and (f). For panels (e)–(g), the $\beta$ values are derived from optically thin MBB models fitted to photometry ranging from 100 to 3000 $\mu$m. All SMG samples are concatenated together and binned into six bins, represented by green hexagons.

centimeter-sized amorphous silicate (a-Sil) grains, while $\beta$ is around 2 for 0.1–10 $\mu$m sized a-Sil grains. This trend aligns with observational findings, where protostellar disks exhibit $\beta \approx 1$ for millimeter-sized dust (Guilloteau et al. 2011), and molecular clouds show $\beta \approx 2$ for micron-sized dust (Sadavoy et al. 2013). In the context of galactic-scale observations, the emission in the Rayleigh–Jeans region mainly originates from submicron- to micron-sized dust grains, as illustrated in Jones et al. (2013). This suggests that the $\beta$ value we found in our sample is less dependent on dust grain size. However, the composition of the dust grains plays a more significant role in determining $\beta$. Simulations conducted by Hirashita et al. (2014) indicate that a $\beta$ value of 2 can result from emissions by either graphite or silicate grains. Experimental studies, such as Inoue et al. (2020), have demonstrated significant variations in $\beta$ across different materials. The issue of $\beta$ and composition does not have clear-cut conclusions, and there is a degeneracy between $\beta$ and the exact dust grain composition. Therefore, in our findings of $\beta \approx 2$ for SMGs at cosmic noon, we cannot definitively identify the dust grain composition. However, because there is no correlation between $\beta$ and redshift shown in

Figure 9(f), it is possible that the dust grain composition may not undergo strong evolution.

To gain a comprehensive understanding of dust grain properties, relying solely on $\beta$ is inherently limited. While $\beta$ provides some insights, it primarily informs us about dust emission in the Rayleigh–Jeans region and allows us to rule out certain dust materials based on their associated $\beta$ values. To conduct a more comprehensive study of dust properties, it is crucial to incorporate the mid-infrared data. In this regard, the JWST emerges as a powerful tool that can significantly contribute to our understanding of this issue. The mid-infrared observations enabled by JWST can provide valuable information about smaller dust grains and PAHs (Spilker et al. 2023), facilitating further investigations into the intricacies of dust grain properties.

### 4.5. Mass Function

Since our sample is a result of a uniform $S_{870}$-selected survey, we can derive mass functions based on our measurements and compare to the model predictions.

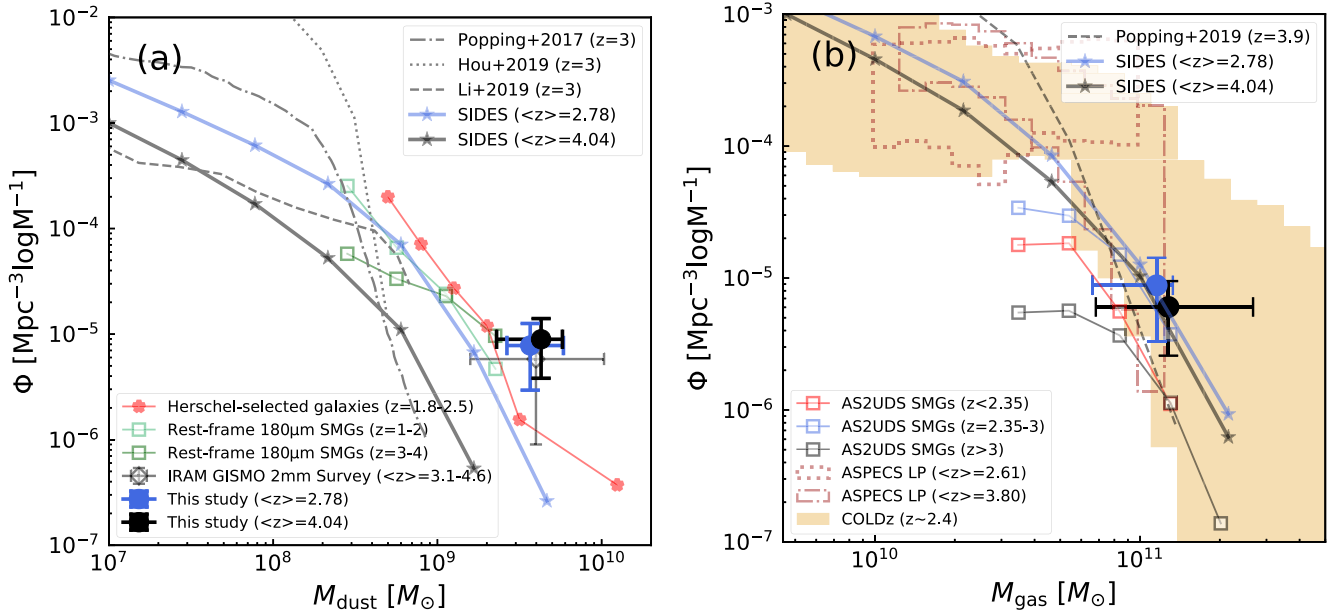

**Figure 10.** (a) Dust mass function derived from our AS2COSPEC SMGs, the IRAM GISMO 2 mm survey sources (Magnelli et al. [2019]), the Herschel-selected galaxies (Pozzi et al. [2019]), and the rest-frame ~180 μm–selected SMGs (Dudzevičiūtė et al. [2021]). The x-axis error bars demonstrate the $M_{dust}$ bin width, while the y-axis error bars are the combination of Poisson errors and the uncertainties from completeness. Predictions from simulations (Popping et al. [2017]; Hou et al. [2019]; Li et al. [2019]; Béthermin et al. [2022]) are plotted in different line styles with huge variances in the low-mass region. (b) Gas mass function derived from our AS2COSPEC SMGs, AS2UDS SMGs (Dudzevičiūtė et al. [2020]), the VLA COLDz survey (Riechers et al. [2019]), and the ASPECS LP (Decarli et al. [2019]). The x-axis error bars demonstrate the $M_{gas}$ bin width, while the y-axis error bars are the combination of Poisson errors and the uncertainties from completeness. Prediction of the semianalytical model is obtained from Popping et al. [2019], and those of the SIDES model (Béthermin et al. [2022]) are converted from their $L'_{CO(1-0)}$ function.

### 4.5.1. Dust Mass Function

We derived the dust mass function at a given redshift bin by

$$\Phi_{M_{dust}} = \sum_{i=1}^{N_i} \frac{1}{\Delta V \Delta M} \times \frac{1}{C}, \quad (12)$$

where $i$ represents the index of sources in a given redshift bin. We divided our sample into two redshift bins: a lower redshift bin with eight sources and a higher redshift bin with nine sources. The redshift bin boundaries are given as $z = [2.26, 3.29, 4.78]$, with the centers of each bin located at $z = 2.78$ and 4.04. $N_i$ represents the number of galaxies within a certain redshift bin, $\Delta V$ corresponds to the comoving volume of the corresponding redshift bin, $\Delta M$ denotes the dust mass bin width for sources within the redshift bin, and $C$ stands for the completeness. Table [9] presents the estimated dust mass function values for the lower and higher redshift bins.

Since our sample is flux-limited, it is important to account for the incompleteness of our selection in mass in order to include all sources within the given dust mass bins. As illustrated in the $M_{dust}$–$S_{870}$ plot provided in Dudzevičiūtė et al. [2020], while the correlation is strong, there is still an ~0.4 dex dispersion in $M_{dust}$ for a given flux. Consequently, our flux-limited selection would miss a portion of likely warm and high $M_{dust}$ sources. To estimate the completeness in mass, for each redshift bin, we first compute the number of sources that are supposed to be above the measured minimum dust mass of our AS2COSPEC SMGs, based on the $M_{dust}$–$S_{870}$ correlation (Dudzevičiūtė et al. [2020]), scaled by the systematic offset between X-CIGALE and MAGPHYS, and the AS2COSMOS number counts (Simpson et al. [2020]). The completeness is then the ratio between the total number of our sample SMGs and the total number of sources above this dust mass limit. As a result, we determine the completeness and Poisson uncertainties for the lower and higher redshift bins to be $15^{+8}_{-6}\%$ and $9^{+4}_{-3}\%$. We then apply this correction factor to adjust the gas mass function.

In Figure [10](a), we compare our estimates to literature data that cover similar redshift ranges. Our measurements align well with the IRAM GISMO 2 mm survey sources from Magnelli et al. [2019], who modeled the dust mass of galaxies observed by the IRAM GISMO instrument by fitting mid-infrared to millimeter photometry using the Draine & Li [2007] model. Furthermore, we include measurements of the Herschel-selected galaxies (Pozzi et al. [2019]) and the rest-frame ~180 μm–selected SMGs from STUDIES and AS2UDS (Dudzevičiūtė et al. [2021]), where the dust mass estimations are derived through MBB fitting to the FIR and millimeter photometry. For the former, the dust temperature $T_{dust}$ was

**Table 9**
Mass Function

|  | $\langle z \rangle = 2.78$ | $\langle z \rangle = 4.04$ |
|---|---|---|
| Completeness [%] | $15^{+8}_{-6}$ | $9^{+4}_{-3}$ |
| $\langle M_{dust}/M_{\odot} \rangle$ | $3.7 \times 10^9$ | $4.3 \times 10^9$ |
| $\Delta M_{dust}$ [$M_{\odot}$] | $3.2 \times 10^9$ | $3.5 \times 10^9$ |
| $\Phi_{M_{dust}}$ | 2.98, 7.79, 12.64 | 3.84, 8.93, 14.05 |
| $\langle M_{gas}/M_{\odot} \rangle$ | $11.5 \times 10^{11}$ | $12.8 \times 10^{11}$ |
| $\Delta M_{gas}$ [$M_{\odot}$] | $6.7 \times 10^{10}$ | $19.9 \times 10^{10}$ |
| $\Phi_{M_{gas}}$ | 3.37, 8.81, 14.28 | 2.59, 6.04, 9.49 |

**Note.** The three values of the mass function shown in each redshift bin represent the lower bound, the value, and the upper bound. Both $\Phi_{M_{gas}}$ and $\Phi_{M_{dust}}$ are in units of $10^{-6}$ Mpc$^{-3}$ dex$^{-1}$.

fixed to the relation in Magnelli et al. (2014), while for the latter, the dust emissivity index $\beta$ was fixed to 1.8.

We also include several simulation predictions for comparison in Figure 10(a). First, Popping et al. (2017) investigates dust evolution using a semianalytical model that incorporates various physical mechanisms of dust production and destruction from $z = 9$ to 0. While their results broadly reproduce the dust mass function in the local Universe, they underestimate the dust mass at high redshift ($z > 1$) in the $M_{dust} \gtrsim 10^{8.5} M_\odot$ range. Second, Hou et al. (2019) employ the GADGET-3 code, an $N$-body hydrodynamic simulation model coupled with a dust-enrichment model to study dust content on a cosmological scale from $z = 5$ to 0. Their simulation fails to reproduce the extended tails observed at the massive end for all redshifts, which are attributed to astration (see Section 3.5 of Hou et al. 2019 for a more detailed discussion). Third, utilizing the SIMBA cosmological hydrodynamic simulation, Li et al. (2019) investigate dust evolution from $z = 6$ to 0. While their predictions show improvement compared to other simulations at $M_{dust} \gtrsim 10^{8.5} M_\odot$, they still fall short of reproducing the most massive end ($M_{dust} \gtrsim 10^9 M_\odot$), which could be caused by too-strong feedback implemented in the SIMBA simulations, so either dust growth is too slow or dust destruction is too efficient. Lastly, we consider the empirical model SIDES (Béthermin et al. 2022), which is shown in Section 4.4.1 to be more successful in reproducing the observed dust-to-stellar ratio. We convert the intensity of the radiation field $\langle U \rangle$ provided in the SIDES catalog to $M_{dust}$ following Draine & Li (2007) and compute the dust mass functions at the two redshift bins. While the SIDES predictions are in broad agreement with the observed values at $z \sim 3$, the observed number density of galaxies with high dust masses exceeds the predicted number at $z \sim 4$, again suggesting the limited predicting power of SIDES at $z \gtrsim 4$.

### 4.5.2. Gas Mass Function

Following the method described in Section 4.5.1, we also derive the gas mass function and present the corresponding values in Table 9. Figure 10(b) presents the gas mass function of our sample divided into two redshift bins, revealing a weak evolution for gas mass function. We compare our estimates with those derived from AS2UDS SMGs presented in Dudzevičiūtė et al. (2020), which are scaled by a factor of 100/32 to the same gas-to-dust ratio as ours and corrected for the dust mass systematic difference between X-CIGALE and MAGPHYS. Our estimations are broadly consistent with the measurements of Dudzevičiūtė et al. (2020) within the uncertainties. We also include results from the VLA COLDz survey (Riechers et al. 2019) and the ASPECS LP (Decarli et al. 2019) by employing $\alpha_{CO} = 0.8$ to convert their $L'_{CO(1-0)}$ into $M_{gas}$. Our estimates also agree with the COLDz survey and the high-$z$ results of the ASPECS (Decarli et al. 2019).

The model predictions from different simulations are also plotted in Figure 10(b). For SIDES, we extract their simulated data and calculate gas mass functions using the same redshift bins as in our analysis, based on the SIDES catalog. The gas mass is calculated by converting $L'_{CO(1-0)}$ to $M_{gas}$ using $\alpha_{CO} = 0.8$. The results from SIDES successfully describe the gas mass function across the mass range. Additionally, we plot the prediction from a semianalytical simulation (Popping et al. 2019) in Figure 10(b). The detailed introduction to the simulation is presented in Somerville et al. (2015) and Popping et al. (2019). The model is broadly consistent with the observational results, which are also argued and discussed by Popping et al. (2019).

## 5. Summary

In this study, we present the results of the AS2COSPEC survey, which is an ALMA band 3 blind CO survey targeting the brightest SMGs selected from the AS2COSMOS sample (Simpson et al. 2020). Building upon the initial results that included line extraction, redshift measurements, and the investigation of lensing effects as presented in Chen et al. (2022b), we extend our analyses to explore the ISM properties in order to gain a better understanding of the underlying physical mechanisms in these 18 primary brightest SMGs at $z = 2$–5. Given a complete selection of these brightest SMGs, our study yields the following key findings:

1.  We conducted a comprehensive analysis of the line properties, including moment calculations, line width, and line luminosity. Using the $r_{31}$ parameter (Birkin et al. 2021; Frias Castillo et al. 2023) and adopting $\alpha_{CO} = 0.8 \pm 0.6$ (Rivera et al. 2018), we find a median line width in FWHM of $610 \pm 50$ km s$^{-1}$ and a median molecular gas mass of $M_{gas} = (\alpha_{CO}/0.8)(1.2 \pm 0.1) \times 10^{11} M_\odot$.

2.  Utilizing a wealth of ancillary multiwavelength data and spectroscopic redshift measurements from Chen et al. (2022b), we employ the multiwavelength SED fitting using X-CIGALE and MBB fitting to characterize the cold dust emission. Our analyses reveal a median total infrared luminosity of $L_{IR} = (1.3 \pm 0.1) \times 10^{13} L_\odot$ and a median stellar mass of $M_* = (1.4 \pm 0.4) \times 10^{11} M_\odot$, suggesting that AS2COSPEC SMGs are galaxies that are some of the most massive and most active in star formation at $z = 2$–5.

3.  The SED modeling analysis indicates that the FIR energy of all AS2COSPEC SMGs is predominantly attributed to dust emission, with most of the AGN-to-total infrared luminosity ratio (fracAGN) being consistent with zero. Besides, through cross-matching with the Chandra COSMOS-Legacy survey (Civano et al. 2016), we obtained X-ray counterparts in four sources, corresponding to an X-ray detection rate of approximately 20%. In addition, only two of our sources show excess in radio emission. These suggest that AGN contribute a minor part of the energy output in the AS2COSPEC sample. However, future MIR observations by the JWST should allow us to improve the constraints with regard to the obscured AGN.

4.  We observe a $1.8 \pm 0.7$ times higher gas fraction ($\mu_{gas}$) in the AS2COSPEC SMGs compared to the fainter SMGs, indicating a larger gas reservoir available for star formation for the brighter ones. The median gas depletion timescale ($t_{depl}$) for our sample is $90 \pm 10$ Myr, which is 25% lower than that of the fainter SMGs. These suggest that the higher SFRs in the AS2COSPEC SMGs could be due to both a larger amount of gas fuel and an increase of efficiency in star formation. On the other hand, in the context of general galaxy populations, we find that at a fixed redshift, location on the MS, and stellar mass, SMGs have comparable $\mu_{gas}$ but significantly lower $t_{depl}$ compared to the less massive and less active star-forming galaxies predominantly on the MS (Tacconi et al. 2020).

This suggests that SMGs have a cold gas reservoir that can be scaled from typical star-forming galaxies but are much more efficient in forming stellar mass, possibly caused by dynamical interactions.

5. For the AS2COSPEC SMGs, the median dust mass derived from X-CIGALE modeling is $(3.7 \pm 0.5) \times 10^9 \, M_\odot$. Given their stellar masses, AS2COSPEC SMGs have one of the highest dust-to-stellar mass ratios, with a median of $(2.1 \pm 1.0) \times 10^{-2}$, four times more than any other DSFG samples, suggesting that bright SMGs are undergoing a phase of rapid dust mass production. Physically motivated models underpredict the observed values, possibly due to too low of a gas-phase metallicity or too-strong AGN feedback.

6. Due to 850 $\mu$m selection, the dust temperature of our sample is biased toward lower values at a given infrared luminosity, with a median optically thin $T_{dust}$ of $29 \pm 2$ K. The median dust emissivity index $\beta = 2.1 \pm 0.1$ agrees with measurements from some previous studies of less luminous sources (Magnelli et al. 2012; Casey et al. 2021; da Cunha et al. 2021). By combining recent $\beta$ measurements of DSFGs in the literature, we find a lack of correlation of $\beta$ with redshift and infrared luminosity. This may suggest common dust grain compositions for infrared luminous galaxies across a large fraction of cosmic time.

7. Finally, we divide our sources into two redshift bins at $z = 2$–5 and observe no significant evolution in the dust and molecular gas mass functions at the high-mass end. These findings align with measurements from other surveys targeting sources at a similar redshift range. However, physically motivated models tend to underpredict the number density of these massive sources, again possibly suggesting too-strong feedback implemented in these models.

Our analyses provide valuable insights into the fundamental relations and ISM properties of the 18 brightest AS2COSPEC SMGs, contributing to the understanding of the evolutionary studies related to gas, dust, and star formation. These findings promote our knowledge of the unique characteristics of SMGs and their role in galaxy evolution. Future analyses of additional AS2COSPEC sources will be beneficial in validating the scaling relations by expanding the sample size.

## Acknowledgments

We thank the referee for a useful report that has improved the manuscript. We thank Hiroyuki Hirashita for enlightening discussions. C.-L.L. and C.-C.C. acknowledge support from the National Science and Technology Council of Taiwan (NSTC 109-2112-M-001-016-MY3 and 111-2112M-001-045-MY3), as well as Academia Sinica through the Career Development Award (AS-CDA-112-M02). Y.Z. is grateful for support from the National Natural Science Foundation of China (No. 12173079). M.F.C. acknowledges the support of the VIDI research program under project number 639.042.611, which is (partly) financed by the Netherlands Organization for Scientific Research (NWO). The Cosmic Dawn Center (DAWN) is funded by the Danish National Research Foundation under grant No. 140. Y.A. acknowledges support by NSFC grant 12173089. I.S. acknowledges support from STFC (ST/X001075/1). H.U. acknowledges support from JSPS KAKENHI grant (20H01953, 22KK0231). This paper makes use of the following ALMA data: ADS/ JAO.ALMA# 2019.1.01600.S. ALMA is a partnership of ESO (representing its member states), NSF (USA) and NINS (Japan), together with NRC (Canada), MOST and ASIAA (Taiwan), and KASI (Republic of Korea), in cooperation with the Republic of Chile. The Joint ALMA Observatory is operated by ESO, AUI/ NRAO and NAOJ.

## Appendix A
## X-CIGALE and MAGPHYS

In our SED fitting, we adopt X-CIGALE as a tool to deduce the physical parameters of the sample SMGs. However, other SED fitting codes do similar work, so it is useful to test whether some of the parameters are more sensitive to different model assumptions and code implementations.

Here, we compare our X-CIGALE fitting results to those from MAGPHYS, as these two codes are typically used to deduce SMG properties. The comparison samples include our sample SMGs and the fainter SMGs presented in Birkin et al. (2021), where multiwavelength photometry and spectroscopic redshifts are also available. For both samples, we run the same X-CIGALE and MAGPHYS model parameters, and the parameters of MAGPHYS are adopted from Dudzevičiūtė et al. (2020), where they tested their modeling against simulated galaxies from the EAGLE simulations (Crain et al. 2015; Schaye et al. 2015).

Figure 11 shows the results of the comparisons. Overall, the fitting quality is comparable between X-CIGALE and MAGPHYS based on the reduced $\chi^2$. In general, the physical parameters are in good agreement. Some notable differences include that the X-CIGALE-based dust masses are about 30%–50% higher than those based on MAGPHYS, which may be due to the higher $\kappa_0$ adopted in MAGPHYS than that in X-CIGALE. SFRs averaged over 10 Myr are systematically higher in X-CIGALE. This is likely due to the fact that X-CIGALE prefers SFH solutions that have more recent bursts of star formation (Section 3.1.1).

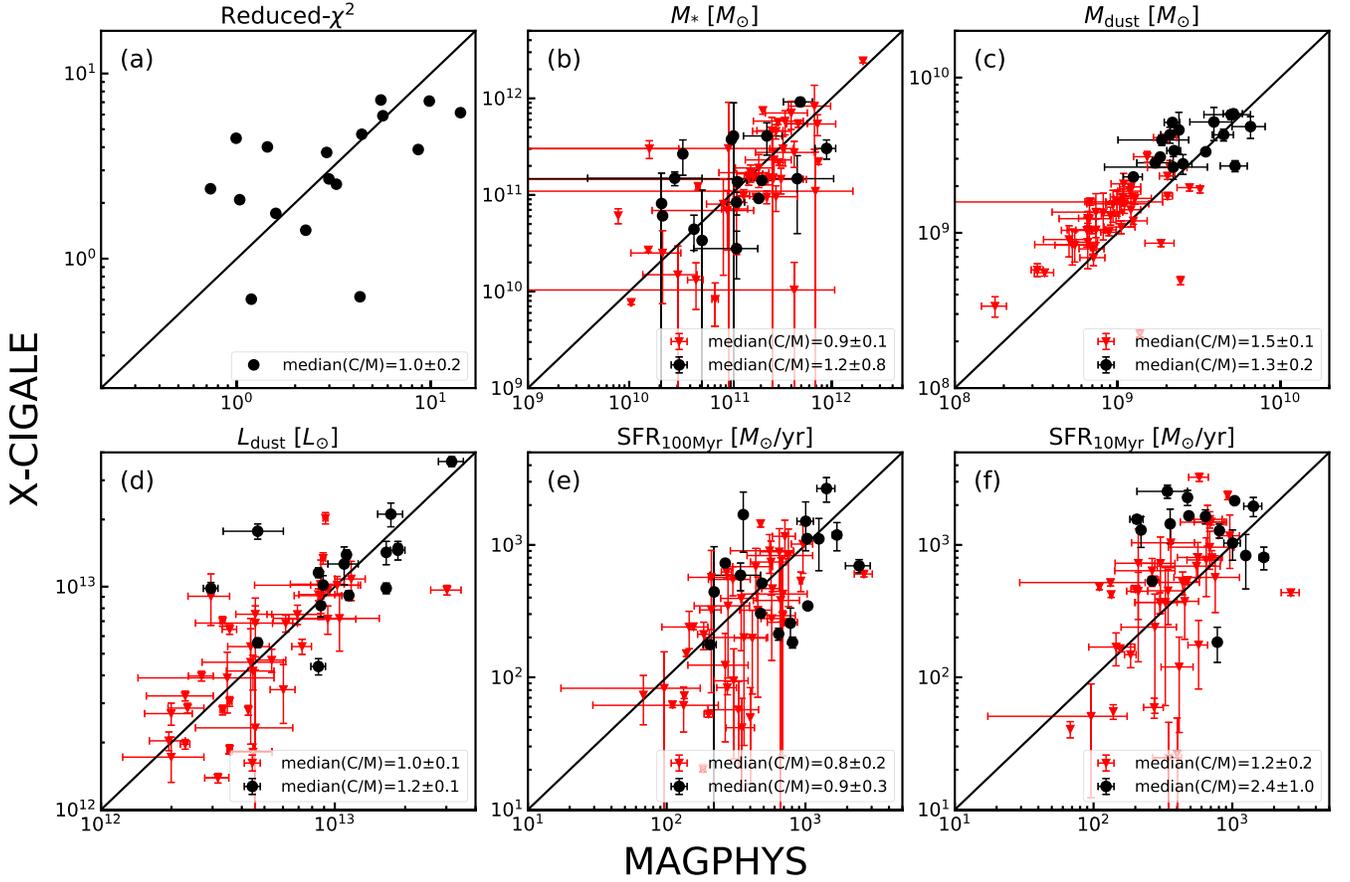

**Figure 11.** Comparison of the SED fitting results from CIGALE and MAGPHYS: (a) the reduced $\chi^2$, (b) the stellar mass, (c) the dust mass, (d) the dust luminosity, (e) the SFR averaged in 100 Myr, and (f) the SFR averaged in 10 Myr. Our SMG sample is plotted with the black dots, and the red triangles demonstrate the typical SMGs in Birkin et al. (2021). Note that for panels (e) and (f), the SFRs of MAGPHYS's estimations are both the 100 Myr averaged values. The medians of the X-CIGALE value to the MAGPHYS value (median(C/M)) for our AS2COSPEC SMGs and sources from Birkin et al. (2021) are shown in the legend in each panel.

## Appendix B
## Best-fit Results of X-CIGALE and MBB

We include all the best-fit X-CIGALE and MBB fitting results of our AS2COSPEC SMGs in Figures 12 and 13.

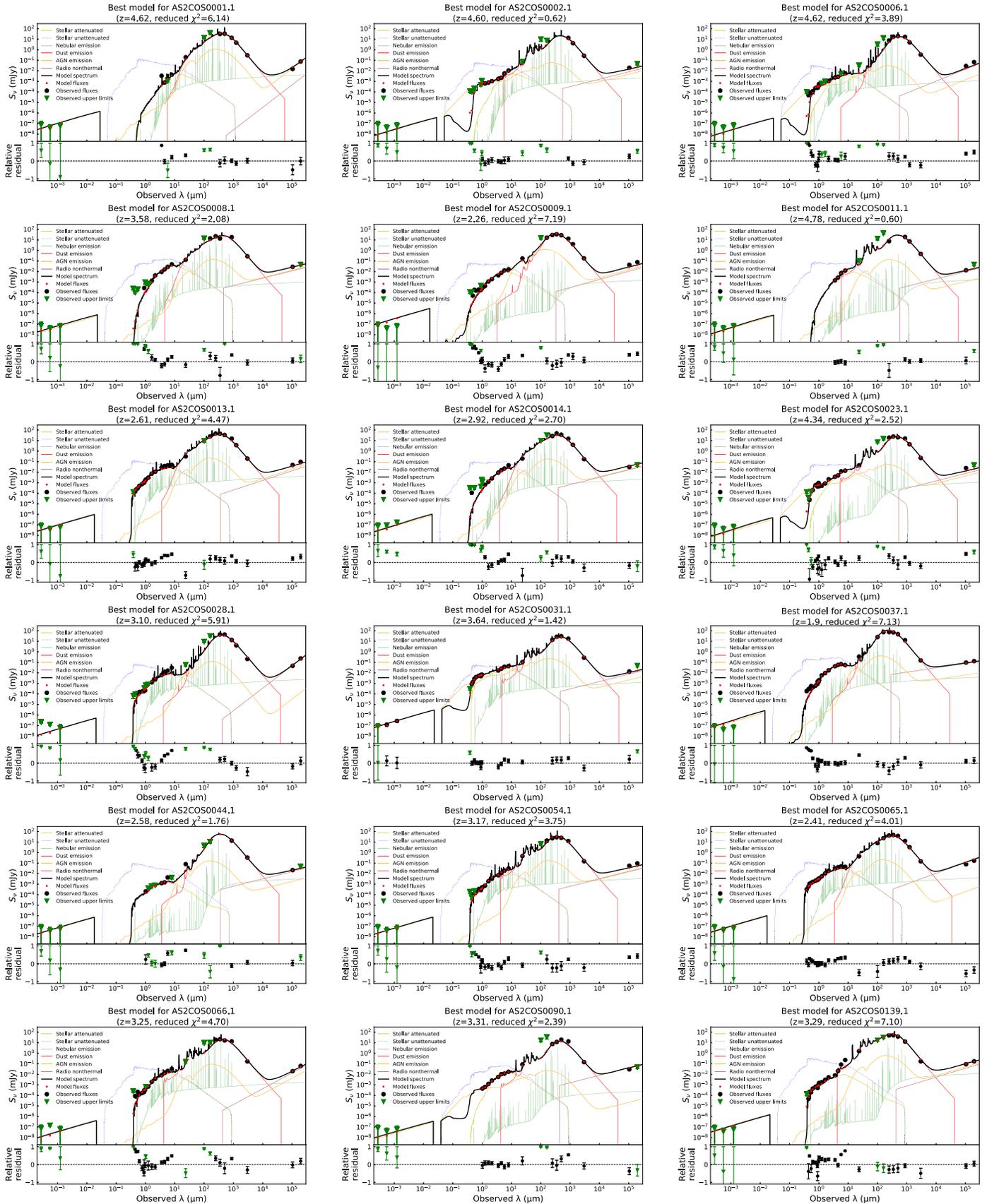

**Figure 12.** Best-fit models of X-CIGALE and MBB for our AS2COSPEC SMGs. Symbols are the same as in Figure 2.

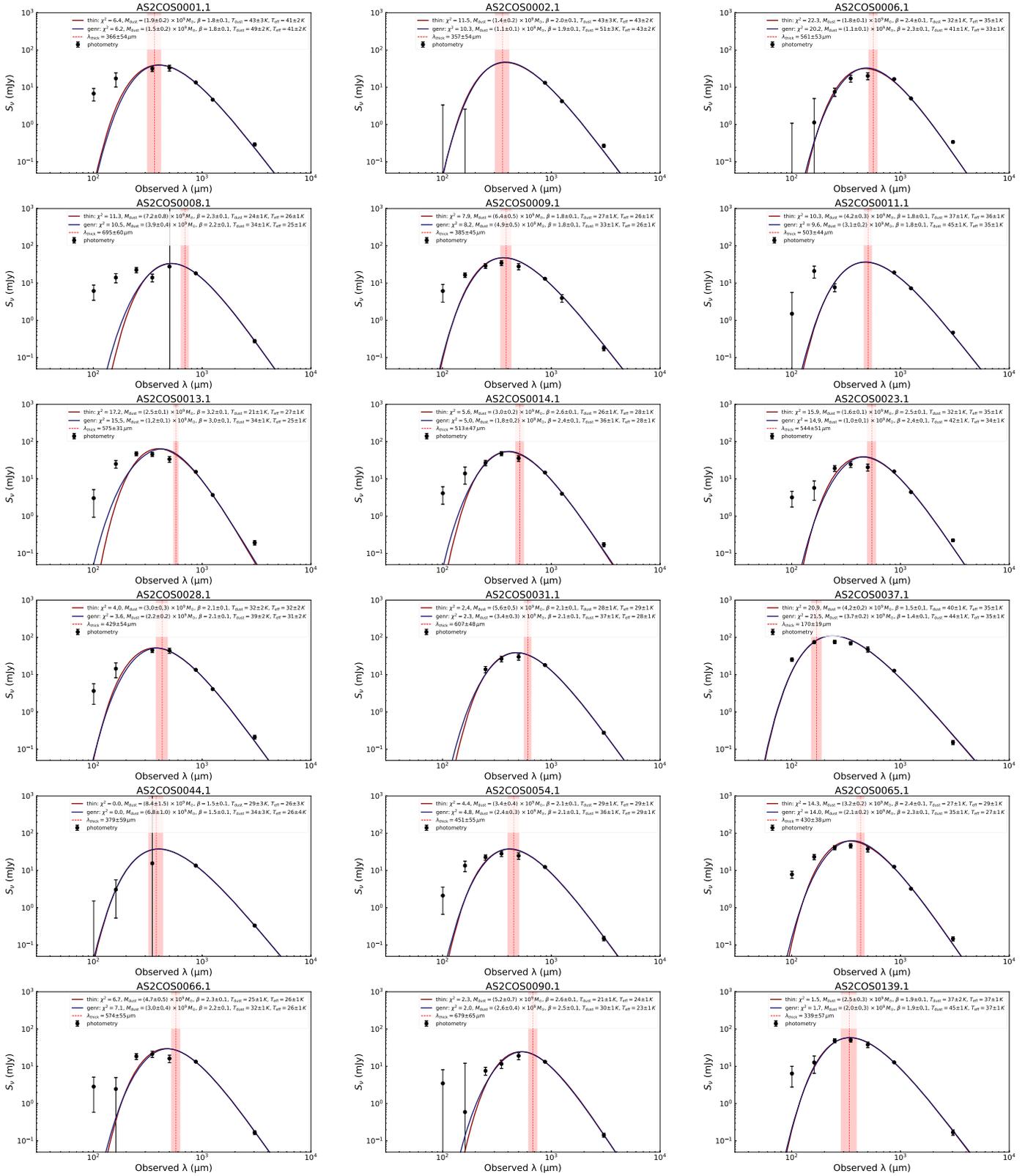

**Figure 13.** Best-fit models of MBB fitting for our AS2COSPEC SMGs. The red vertical line marks the observed-frame $\lambda_{\text{thick}}$, and other symbols are the same as in Figure 2.

## ORCID iDs

Cheng-Lin Liao (廖政霖)  https://orcid.org/0000-0002-5247-6639

Chian-Chou Chen (陳建州)  https://orcid.org/0000-0002-3805-0789

Wei-Hao Wang (王為豪)  https://orcid.org/0000-0003-2588-1265

Ian Smail  https://orcid.org/0000-0003-3037-257X

Y. Ao  https://orcid.org/0000-0003-3139-2724

U. Dudzevičiūtė  https://orcid.org/0000-0003-4748-0681

M. Frias Castillo  https://orcid.org/0000-0002-9278-7028

Minju M. Lee  https://orcid.org/0000-0002-2419-3068

Stephen Serjeant  https://orcid.org/0000-0002-0517-7943

A. M. Swinbank  https://orcid.org/0000-0003-1192-5837

Dominic J. Taylor  https://orcid.org/0000-0002-0031-630X

Hideki Umehata  https://orcid.org/0000-0003-1937-0573

Y. Zhao  https://orcid.org/0000-0002-9128-818X